\documentclass[useAMS,usenatbib,twocolumn]{mn2e}
\usepackage{times,amsmath,amssymb,amsfonts,latexsym}
\usepackage{fleqn,multirow,graphicx,color}
       
\def\mpch{\,{h^{-1} {\rm Mpc}}}          \def\hmpc{\,{h {\rm Mpc}^{-1}}}
\newcommand{\dd}{{\rm d}}                
             \newcommand{\dm}{{\rm dm}}
   \newcommand{\ba}{{\rm b}}
\newcommand{\vv}{{\bf v}}                \newcommand{\vk}{{\bf k}}
\newcommand{\vx}{{\bf x}}                
\newcommand{\apj}{ApJ}                   
\newcommand{\mnras}{MNRAS}               \newcommand{\aap}{A\&A}
\newcommand{\araa}{ARA\&A}               
                     
\newcommand{\nat}{Nature}                
\newcommand{\prl}{Phys. Rev. Lett.}

\begin{document}
\title[Deviation and suppression of baryons from dark matter]{The spatial distribution deviation and the power suppression of baryons from dark matter}
\author[Yang et al.]{Hua-Yu Yang$^1$, Yun Wang$^1$, Ping He$^{1,2}$\thanks{E-mail: hep@jlu.edu.cn},
Weishan Zhu$^3$ and Long-Long Feng$^{3,4}$ \\
$^{1}$College of Physics, Jilin University, Changchun 130012, P.R. China. \\
$^{2}$Center for High Energy Physics, Peking University, Beijing 100871, P.R. China. \\
$^{3}$School of Physics and Astronomy, Sun Yat-Sen University, Guangzhou 510275, P.R. China. \\
$^{4}$Purple Mountain Observatory, CAS, Nanjing 210008, P.R. China.}
\date{\today}
\maketitle
\begin{abstract}
The spatial distribution between dark matter and baryonic matter of the Universe is biased or deviates from each other. In this work, by comparing the results derived from IllustrisTNG and WIGEON simulations, we find that many results obtained from TNG are similar to those from WIGEON data, but differences between the two simulations do exist. For the ratio of density power spectrum between dark matter and baryonic matter, as scales become smaller and smaller, the power spectra for baryons are increasingly suppressed for WIGEON simulations; while for TNG simulations, the suppression stops at $k=15-20\hmpc$, and the power spectrum ratios increase when $k>20\hmpc$. The suppression of power ratio for WIGEON is also redshift-dependent. From $z=1$ to $z=0$, the power ratio decreases from about 70\% to less than 50\% at $k=8\hmpc$. For TNG simulation, the suppression of power ratio is enhanced with decreasing redshifts in the scale range $k>4\hmpc$, but is nearly unchanged with redshifts in $k<4\hmpc$. These results indicate that turbulent heating can also have the consequence to suppress the power ratio between baryons and dark matter. Regarding the power suppression for TNG simulations as the norm, the power suppression by turbulence for WIGEON simulations is roughly estimated to be 45\% at $k=2\hmpc$, and gradually increases to 69\% at $k=8\hmpc$, indicating the impact of turbulence on the cosmic baryons are more significant on small scales.
\end{abstract}

\section{Introduction}
\label{sec:intro}

Both observations and theories of contemporary cosmology reveal that the spatial distribution of the cosmic dark matter and baryons is biased or deviates from each other. In a previous work \citep{Yang2020}, we make use of numerical simulation data produced by the cosmological $N$-body/hydrodynamical code WIGEON \citep{Zhu2013}, to study to what extent the deviation of spatial distribution between dark matter and baryons is. By computing the cross-correlation functions of density field and velocity field for the two matter components, we find that deviations between dark matter and baryonic matter are the most prominent on small scales and diminish gradually on increasingly large scales. The deviations are also time-dependent, becoming increasingly large with cosmic time. The significant result is that deviations of the spatial distribution between the two matter components uncovered by the velocity field are more remarkable than by the density field. In the simulations, we do not include stellar formation and evolution, metal enrichment, growth of black holes, or any baryonic feedback processes, and we attribute the spatial distribution deviation between the two matter components to the turbulent heating to the inter-galactic medium (IGM) or intra-cluster medium (ICM).

Previous theoretical studies also revealed that, at low redshifts, the highly evolved IGM or ICM can be characterized by the She–Leveque's universal scaling formula, suggesting that the cosmic baryon fluids are in a fully developed turbulent state \citep{Zhu2010, Hep2006, Fang2011,  Zhu2015, Zhu2013, Zhu2017}. These results are supports to the idea of turbulent heating to the IGM or ICM.

Generally, turbulence occurring in the fluid depends on the condition that the dimensionless Reynolds number should be sufficiently large. In this case, laminar motions will spontaneously turn into turbulent motions in the fluid. In the context of cosmology, turbulence influences IGM or ICM by providing both thermal and kinetic effects, of which the latter is termed {\em turbulent pressure}, to the cosmic baryonic gas \citep{Bonazzola1987, Bonazzola1992, Zhu2010, Zhu2011}. According to \citet{Bauer2012}, turbulence in IGM or ICM can be categorized into subsonic turbulence and supersonic turbulence. The subsonic turbulence is well described by Kolmogorov theory, in that it yields Kolmogorov-like universal scaling laws for the power spectrum of density, velocity, and vorticity in IGM or ICM, while the supersonic turbulence is characterized by a complex shock-wave web, not described by the Kolmogorov theory but by Burgers turbulence. \citet{Zhu2010} demonstrate that the cosmic baryons are in the fully developed turbulence on scales $<3\mpch$, and they also reveal that the turbulent pressure is roughly equivalent to the thermal pressure of IGM or ICM with the temperature $\sim$$10^6$ K in regions with the mean cosmic baryon density. The turbulent pressure, different from the heating mechanisms by the supernova (SN) or active galactic nucleus (AGN) feedback, is basically non-thermal and dynamical, and hence does not affect the ionizing processes and thermal states of hydrogen in the baryonic gas.

During gravitational collapsing of IGM, both subsonic turbulence and supersonic turbulence may emerge in the cosmic baryon fluids when the Reynolds number is high enough. In the case of subsonic turbulence, the gravitational potential energy of baryon gas is transformed into the kinetic energy, then cascades from large scales to increasingly smaller scales, and eventually dissipates to heat the IGM. In the case of supersonic turbulence, the turbulence can directly heat the IGM through shocks and also provide ram pressure to the IGM. Hence we see that turbulence will contribute both thermal effects and turbulent pressure to the IGM, and in this way, the fully developed turbulence will be a heating mechanism, preventing the IGM from falling into the gravitational potential well of dark halos.

There is a so-called `cooling crisis' or overcooling problem of hierarchical galaxy formation \citep{Voit2005}, and some heating mechanisms proposed to resolve the problem, such as feedbacks like the galactic winds from star formation and the SN explosions \citep{Dekel1986, White1978, White1991}, or AGN activities \citep{Silk1998}, see also the review articles by \citet{Heckman2014} or \citet{Fabian2012}. These feedback processes are powerful heating mechanisms that are able to heat the gas in and around dark matter halos, and prevent the baryon gas from being accreted and forming stars, or expel the gas directly from dark halos \citep{Somerville2015}. Simulations show that the heating mechanisms, particularly AGN feedback, make important contributions to separate baryonic matter from dark matter in the spatial distribution. For example, OWLS \citep{vanDaalen2011}, Illustris \citep{Vogelsberger2014}, EAGLE \citep{Hellwing2016}, BAHAMAS \citep{McCarthy2017, McCarthy2018}, Horizon \citep{Chisari2018}, and IllustrisTNG \citep{Springel2018} show that feedback from AGN can make a remarkable impact on the total density power spectrum. However, there is no consensus in which scale range and to what extent that AGN activities affect the power spectrum of total matter. All the simulations mentioned above reach percent-level deviation at $k\sim1\hmpc$, while OWLS, Illustris and BAHAMAS reach it at even $k\sim0.2 - 0.3\hmpc$, but all the simulations agree that the power spectra are suppressed by about 10 -- 30\% at $k\sim10\hmpc$ \citep{vanDaalen2020}. For more details about the baryonic effects on matter clustering in the context of cosmological hydrodynamical simulations, we refer the readers to a review of \citet{Chisari2019}.

In this work, by comparing the results obtained from IllustrisTNG and WIGEON data, we check whether the conclusions derived in \citet{Yang2020} still hold for IllustrisTNG simulations. If there are some differences between them, what are the reasons for the differences? We organize the paper as follows. Firstly, we introduce the theoretical basis, including the non-linear bias model and the two cross-correlation functions for the deviation between dark matter and baryon matter in Section~\ref{sec:tbasis}. We demonstrate that the bias functions in the Fourier space should be complex functions of the Fourier mode $\vk$. Secondly, we introduce briefly the simulations used in this work in Section~\ref{sec:data}. Then, we give results about the environment- and scale-dependent deviations, and present discussions about the deviation mechanisms of the density and velocity field in Section~\ref{sec:result}. Finally, we summarize our work and present the conclusions in Section~\ref{sec:concl}.

\begin{figure}
\centerline{\includegraphics[width=0.45\textwidth]{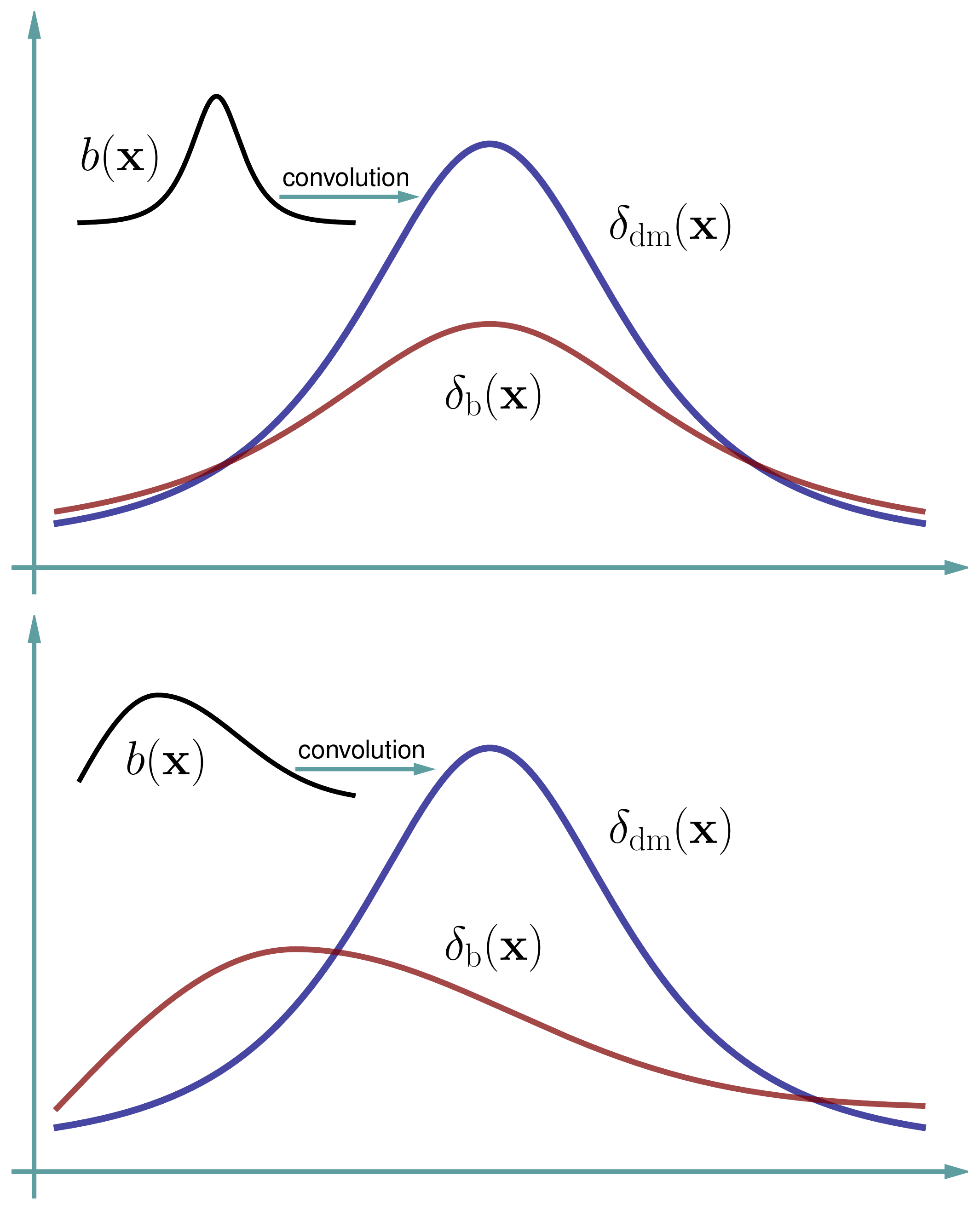}}
\caption{Illustration of the bias model described in Section~\ref{sec:tbasis}. In the upper panel, the distribution of baryons ($\delta_{\ba}(\vx)$) is suppressed from dark matter ($\delta_{\dm}(\vx)$), but not biased. In the lower panel, the distribution of baryons is both suppressed and biased from dark matter.}
\label{fig:bias}
\end{figure}
\begin{figure*}
\centerline{\includegraphics[width=0.85\textwidth]{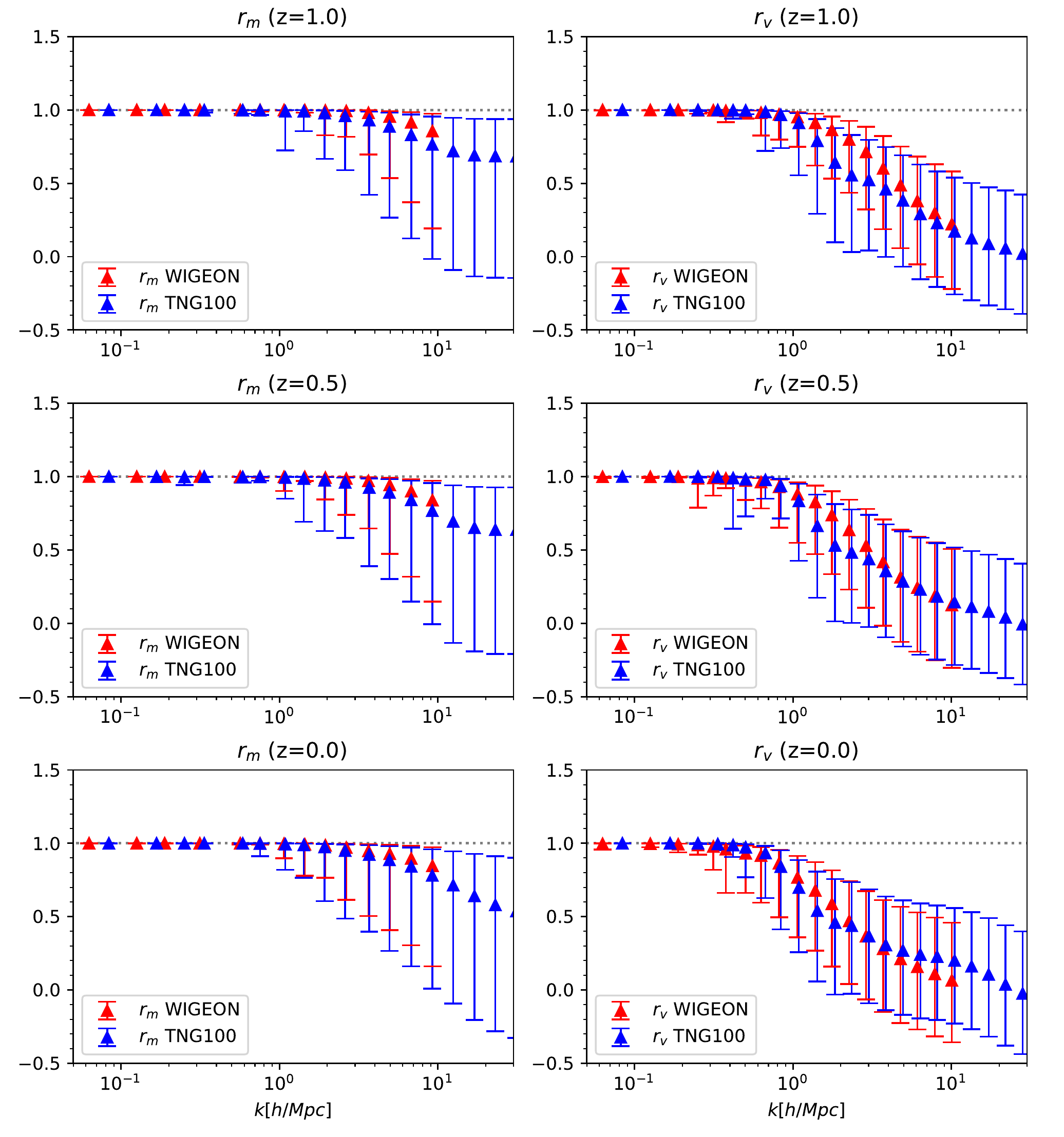}}
\caption{Correlation functions $r_{\rm m}(k)$ (left panels) and $r_{\rm v}(k)$ (right panels) for both WIGEON and TNG100 simulations. Red and blue markers are for $r_{\rm m}$ and $r_{\rm v}$, respectively. Redshifts $z=1$, $0.5$ and $0$ are shown from top to bottom. The valid scale range is restricted to be $k<8\hmpc$ for WIGEON data. One can find the reason of such choice in \citet{Yang2020}. The error bars are computed in the same way as in \citet{Yang2020}.}
\label{fig:rmrv}
\end{figure*}

\section{Theoretical Basis: Cross-correlation Functions}
\label{sec:tbasis}

As demonstrated in \citet{Yang2020}, the linear bias model is oversimplified and is not able to describe the bias of spatial distribution between dark matter and baryon matter. In this work, we continue to use the non-linear bias model previously used as follows,
\begin{equation}
\label{eq:bias}
\delta_{\rm b}(\vk) = b(\vk)\delta_{\rm dm}(\vk),
\end{equation}
where $\vk$ is the Fourier mode, and $b(\vk)$ is the bias function between $\delta_{\rm dm}(\vk)$ and $\delta_{\rm b}(\vk)$. Equation~(\ref{eq:bias}) is still a linear bias form, but if $b(\vk)$ is a complex function, it can be used to describe the non-linear deviation between dark matter and baryons. In Fig.~\ref{fig:bias}, we show the picture for this bias mechanism.

In order to investigate the spatial-distribution deviation between dark matter and baryons, we construct the following two cross-correlation functions, $r_{\rm m}(k)$ and $r_{\rm v}(k)$, which are scale dependent and hence are functions of $k$, for the density field (`m') and the velocity field (`v'), respectively, as \citep{Yang2020},
\begin{eqnarray}
\label{eq:rmvk}
r_{\rm m}(k) & = & \langle\frac{\delta_{\rm dm}(\vk)\delta^{*}_{\rm b}(\vk)}{|\delta_{\rm dm}(\vk)|
|\delta_{\rm b}(\vk)|}\rangle, \nonumber \\
r_{\rm v}(k) & = & \langle\frac{\vv_{\rm dm}(\vk)\cdot\vv^*_{\rm b}(\vk)}{|\vv_{\rm dm}(\vk)|
|\vv_{\rm b}(\vk)|}\rangle,
\end{eqnarray}
where `$<...>$' denotes the statistical average among all the modes with the modulus $k$, and `$*$' denotes the operation of complex conjugate. The bias model equation~(\ref{eq:bias}) is related to $r_{\rm m}$ or $r_{\rm v}$ coefficients as follows. If $b(\vk)$ is not a real, but a complex function of the Fourier mode $\vk$, i.e. $b(\vk) = |b(\vk)|(\cos\theta_b(\vk) + i \sin\theta_b(\vk))$, then
\begin{eqnarray}
\label{eq:rcmplx}
r_{\rm m}(k) & = & \langle\frac{\delta_{\rm dm}(\vk)\delta^*_{\rm b}(\vk)}{|\delta_{\rm dm}(\vk)|
|\delta_{\rm b}(\vk)|}\rangle = \langle\frac{b^*(\vk)}{|b(\vk)|}\rangle \nonumber \\
& = & \langle\cos\theta_b(\vk)\rangle - i \langle\sin\theta_b(\vk)\rangle.
\end{eqnarray}
Notice that if the spatial-distribution deviation between dark matter and baryons is characterized by linear bias model, where $b(\vk)$ is real, then whatever the form of the bias function is, $r_{\rm m}(k)$ is always equal to one \citep{Yang2020}. We can also obtain similar results for $r_{\rm v}(k)$. Hence the two correlation functions, which are defined in equation~(\ref{eq:rmvk}), entirely reflect the non-linear deviations in the spatial distribution between dark matter and baryons.

Actually, $r_{\rm m}(k)$ (or $r_{\rm v}(k)$) is just the average of cosine of $b(\vk)$'s argument $\theta_b(\vk)$, as $r_{\rm m}(k) = \langle\cos \theta_b(\vk)\rangle$, since the `sine' term in equation~(\ref{eq:rcmplx}) is statistically vanishing. Furthermore, we make a rough assumption that $\theta_b(\vk)$ distributes uniformly between $-\theta_0$ and $+\theta_0$, and hence $r_{\rm m}(k)$ is estimated as
\begin{equation}
\label{eq:rmest}
r_{\rm m}(k) = \langle\cos\theta_b(\vk)\rangle \simeq \frac{1}{2\theta_0} \int^{ \theta_0 }_{ -\theta_0} \cos\theta_b(\vk)\dd\theta_b(\vk) = \frac{\sin\theta_0}{\theta_0}.
\end{equation}
If the scatter of $\theta_b(\vk)$ distribution is small, i.e. small $\theta_0$, then $r_{\rm m}(k)$ will approach one, as is observed in Fig.~\ref{fig:rmrv} when $k \rightarrow 0$; while if the scatter is large, then $r_{\rm m}(k)$ will become smaller than one, as is the case for large $k$.

Incidentally, it would be interesting if one can present an analytic or semi-analytic form for $b(\vk)$ in equation~(\ref{eq:bias}), which can properly describe its dependence on $\vk$ and its time evolution. This analytic/semi-analytic form may be useful for modelling of the spatial distribution of the baryonic matter.

\begin{figure}
\centerline{\includegraphics[width=0.5\textwidth]{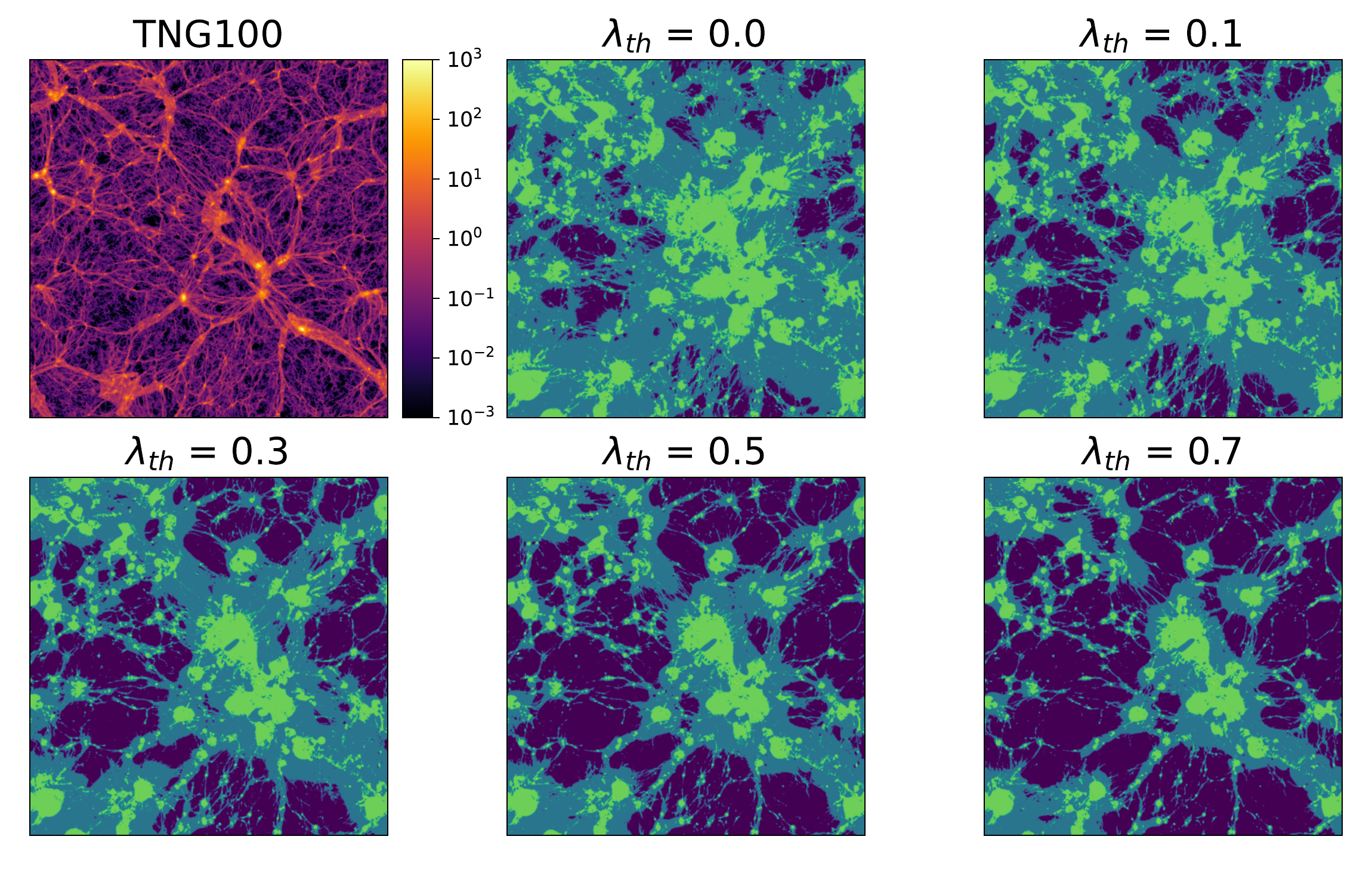}}
\centerline{\includegraphics[width=0.5\textwidth]{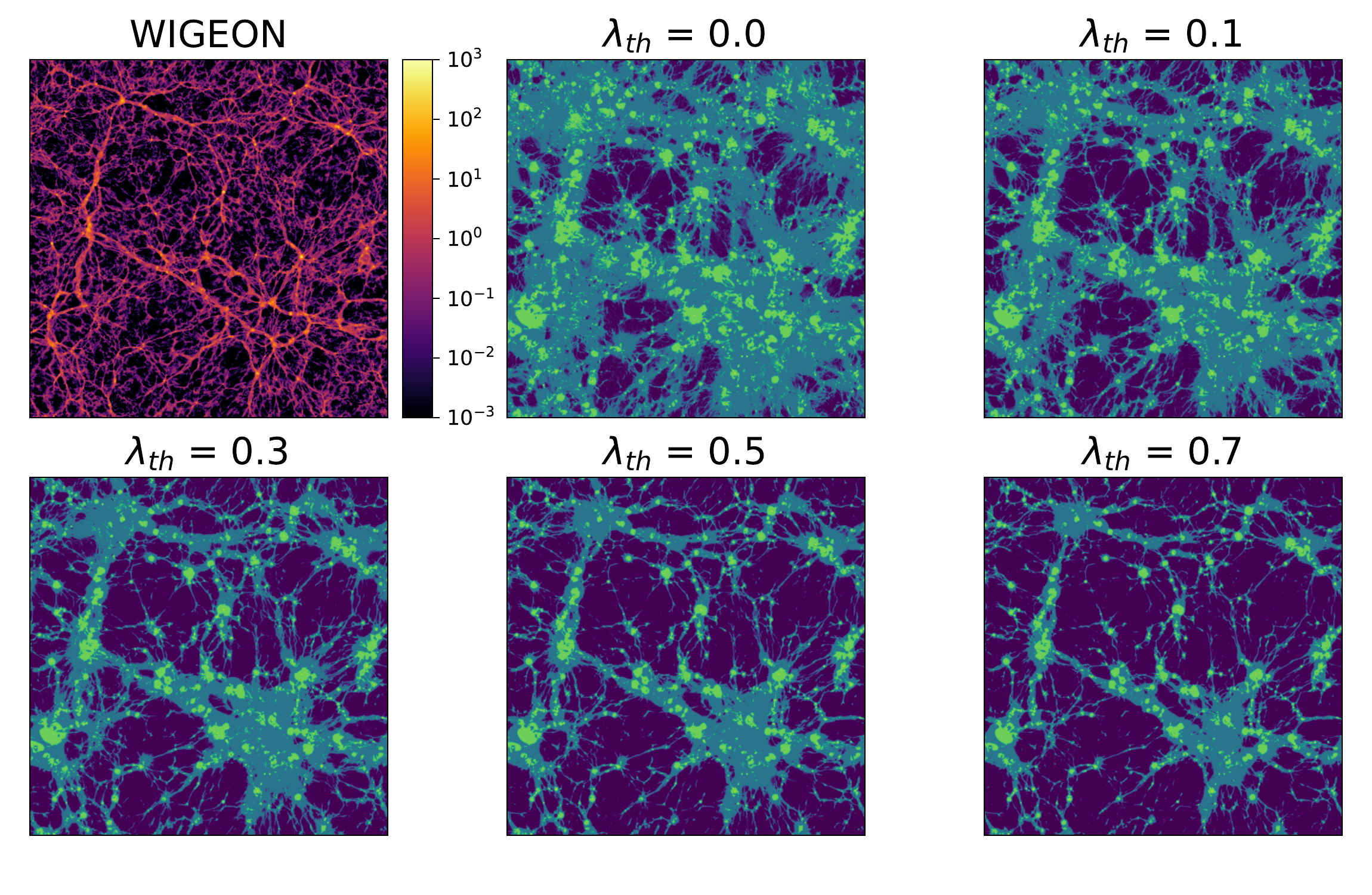}}
\caption{The two panels with color scales are two-dimensional slice images of the cosmic web, to compare with the corresponding spatial distribution of dark matter with different $\lambda_{\rm th}$. The upper two rows are produced by TNG100 data, and the lower two rows by WIGEON data. In the $\lambda_{\rm th}$ panels, the colors from the dark-most to the bright-most indicate voids, sheets, filaments, and clusters, respectively. All data are taken at $z=0$.}
\label{fig:lambda-th}
\end{figure}

\begin{figure*}
\centerline{
\includegraphics[width=0.40\textwidth]{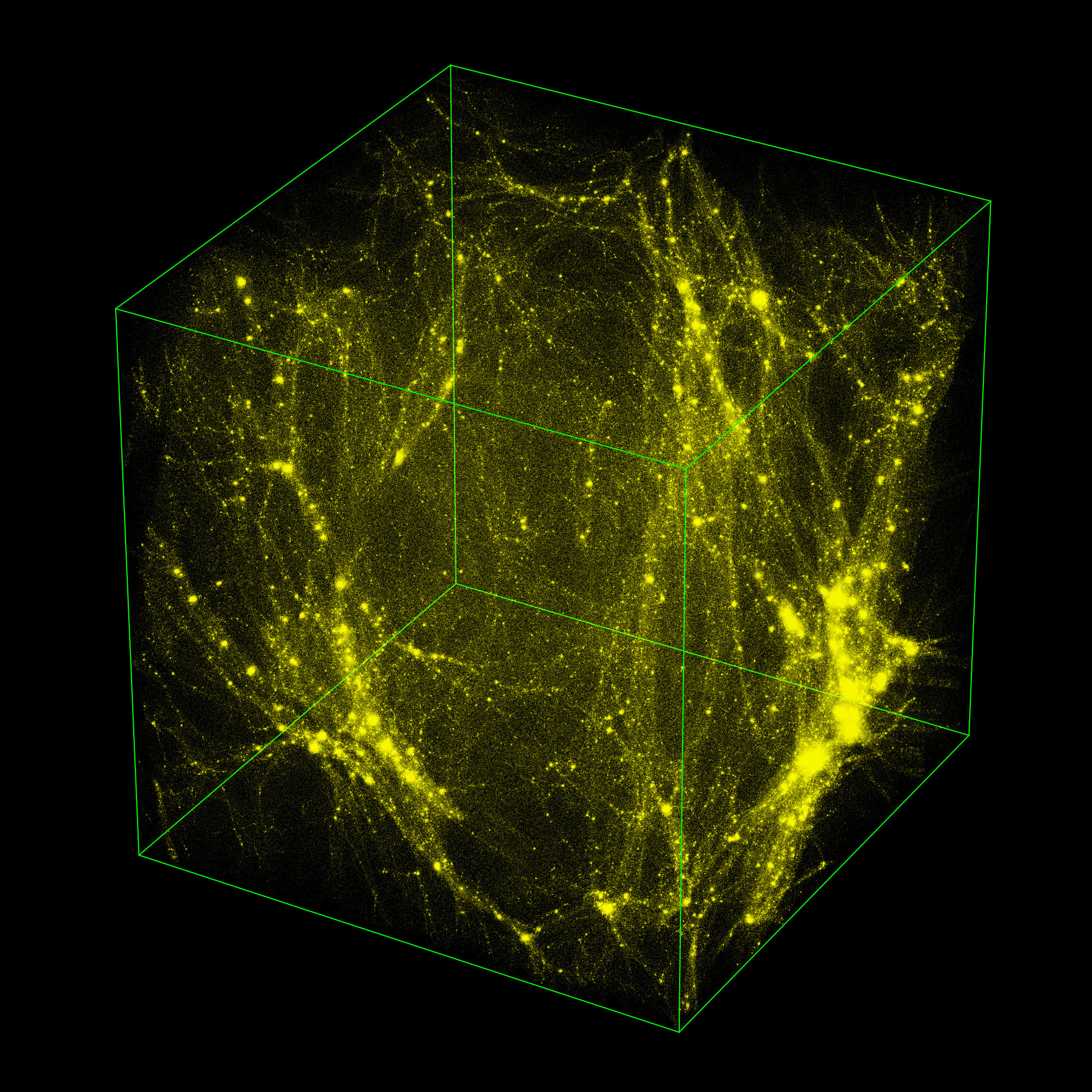} {\hskip 10pt}
\includegraphics[width=0.40\textwidth]{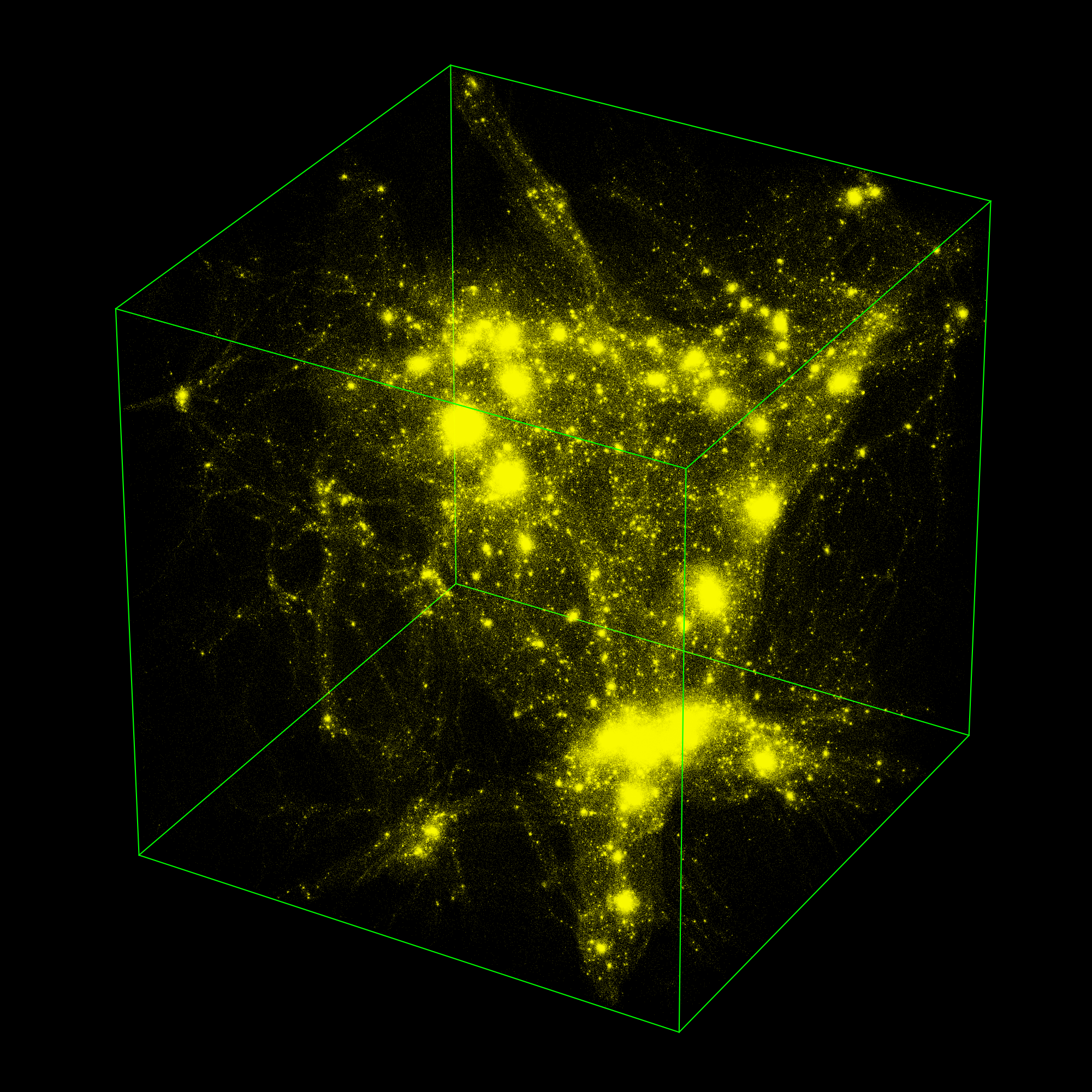}
} {\vskip 10pt}
\centerline{
\includegraphics[width=0.40\textwidth]{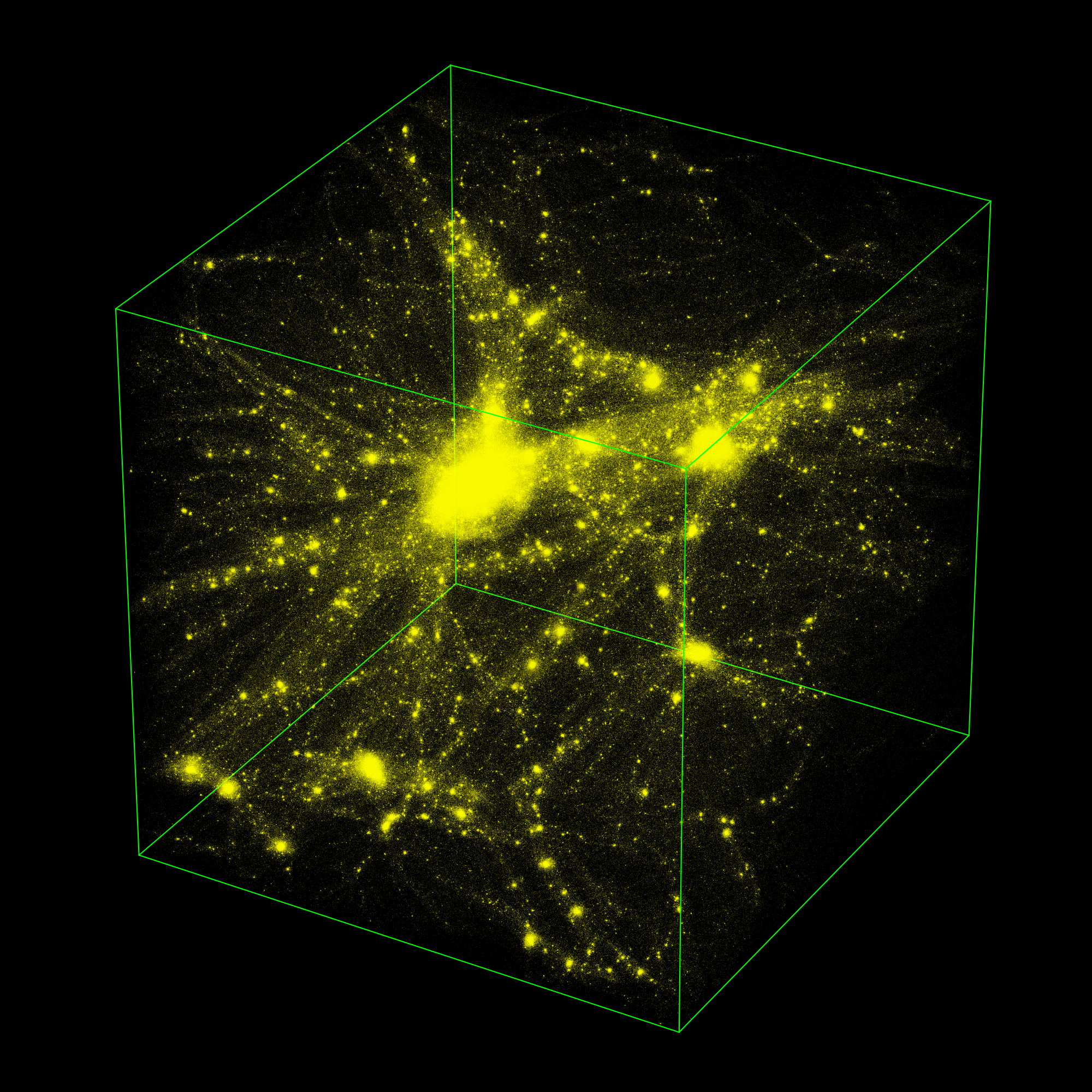} {\hskip 10pt}
\includegraphics[width=0.40\textwidth]{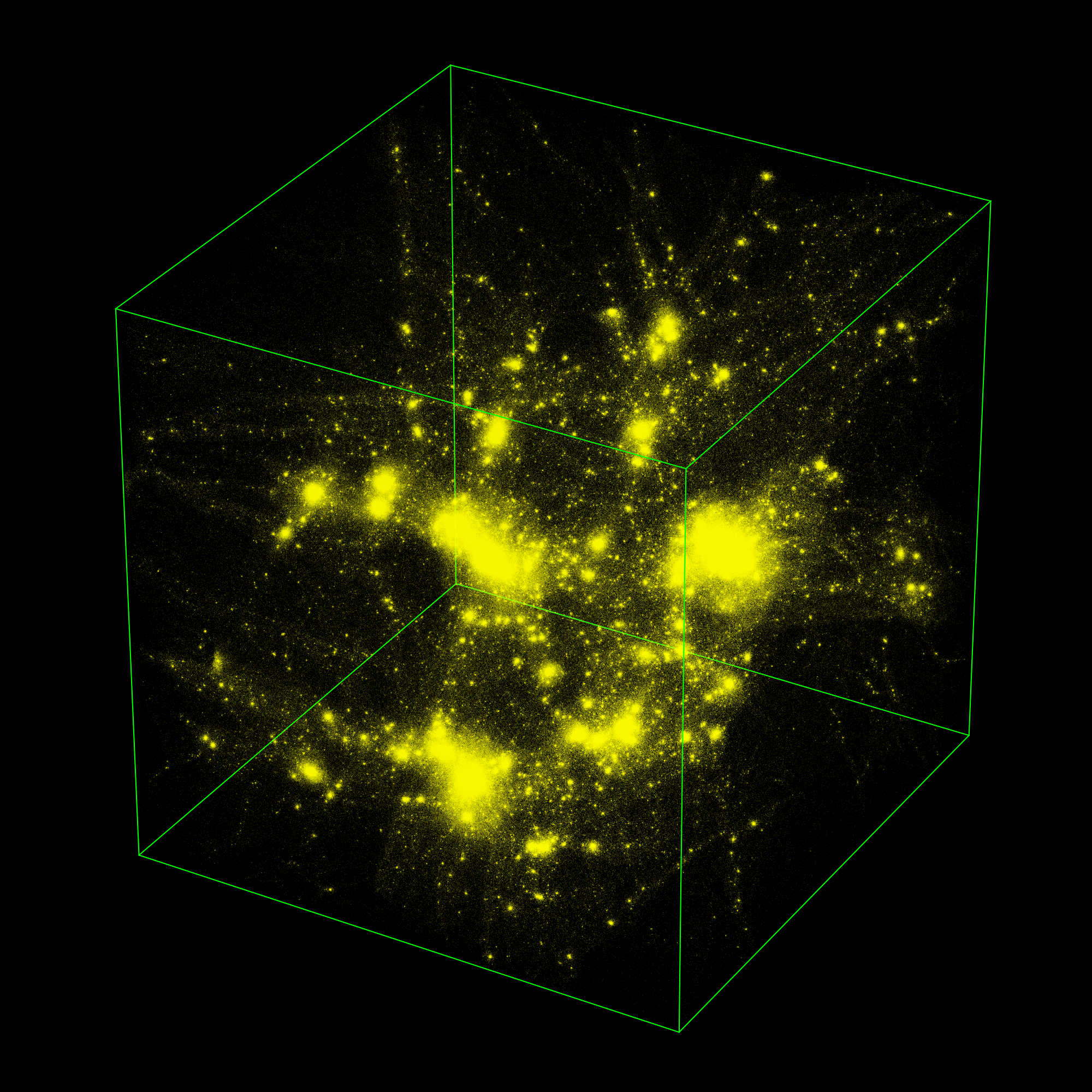}
}
\caption{Three-dimensional views of the four cosmic structures at $z=0$, selected from TNG100 data. The views are for a void (top-left), a sheet (top-right), a filament (bottom-left), and a cluster (bottom-right), respectively. The length of each side of the box is $14.06\mpch$.}
\label{fig:tng4z00}
\end{figure*}

\begin{figure*}
\centerline{
\includegraphics[width=0.40\textwidth]{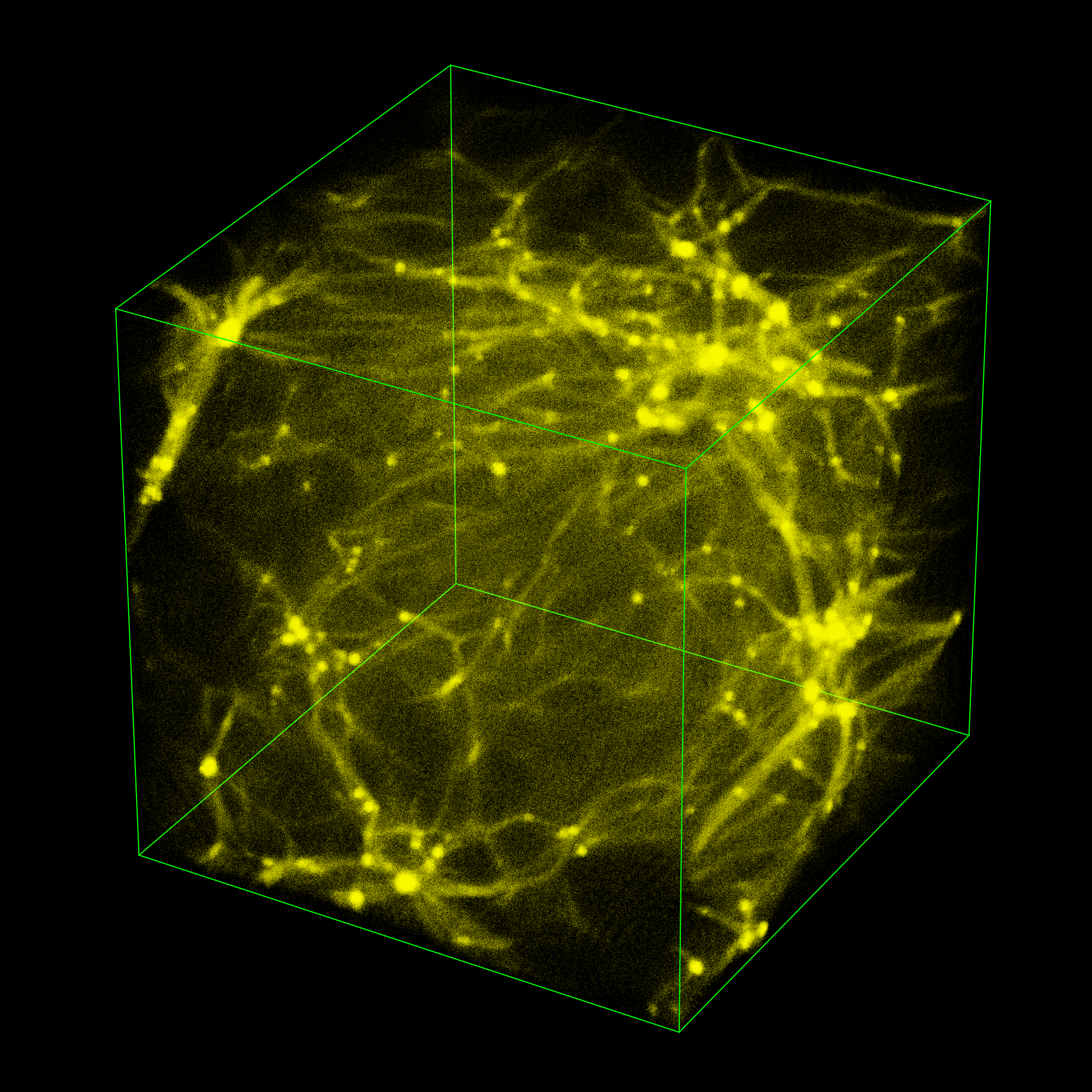} {\hskip 10pt}
\includegraphics[width=0.40\textwidth]{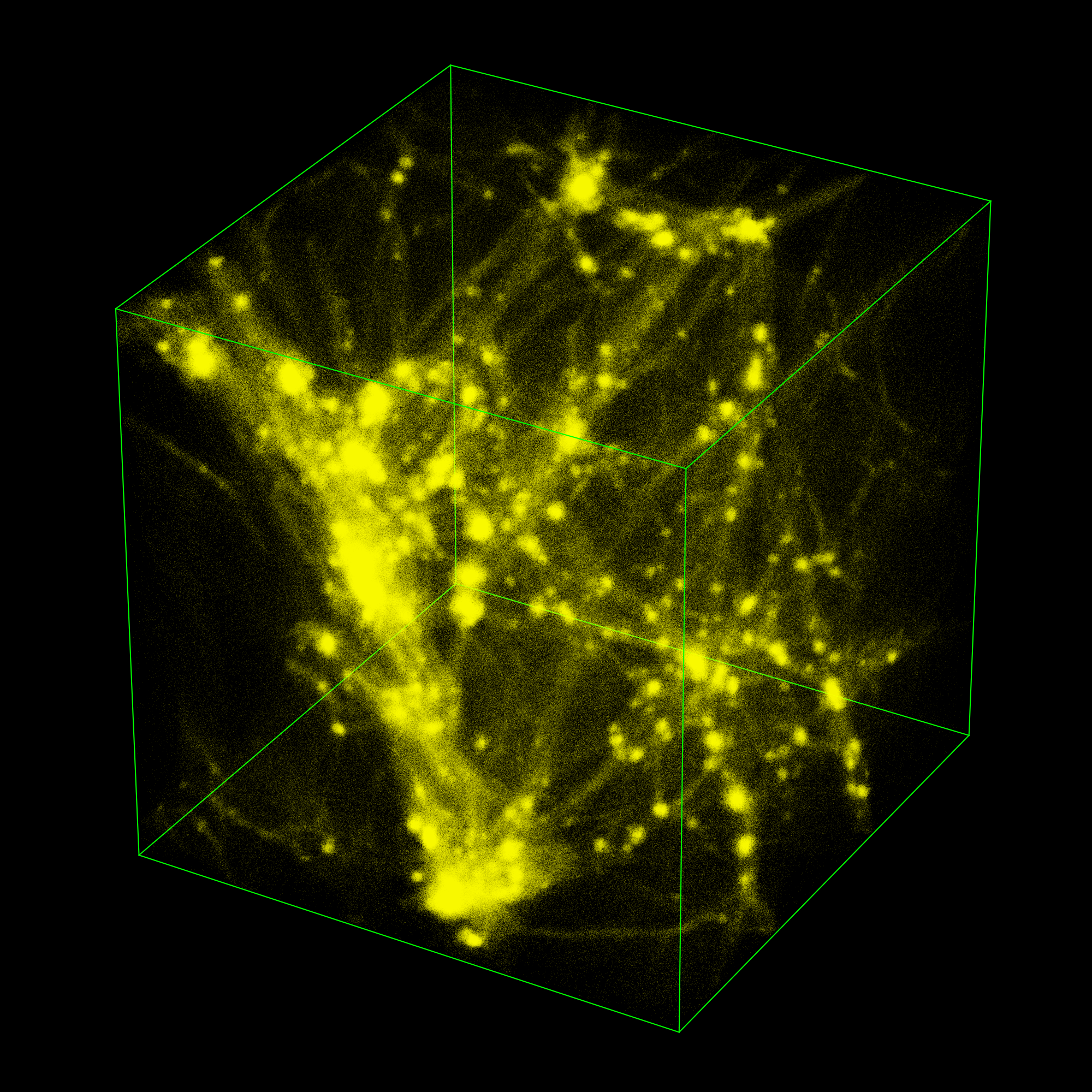}
} {\vskip 10pt}
\centerline{
\includegraphics[width=0.40\textwidth]{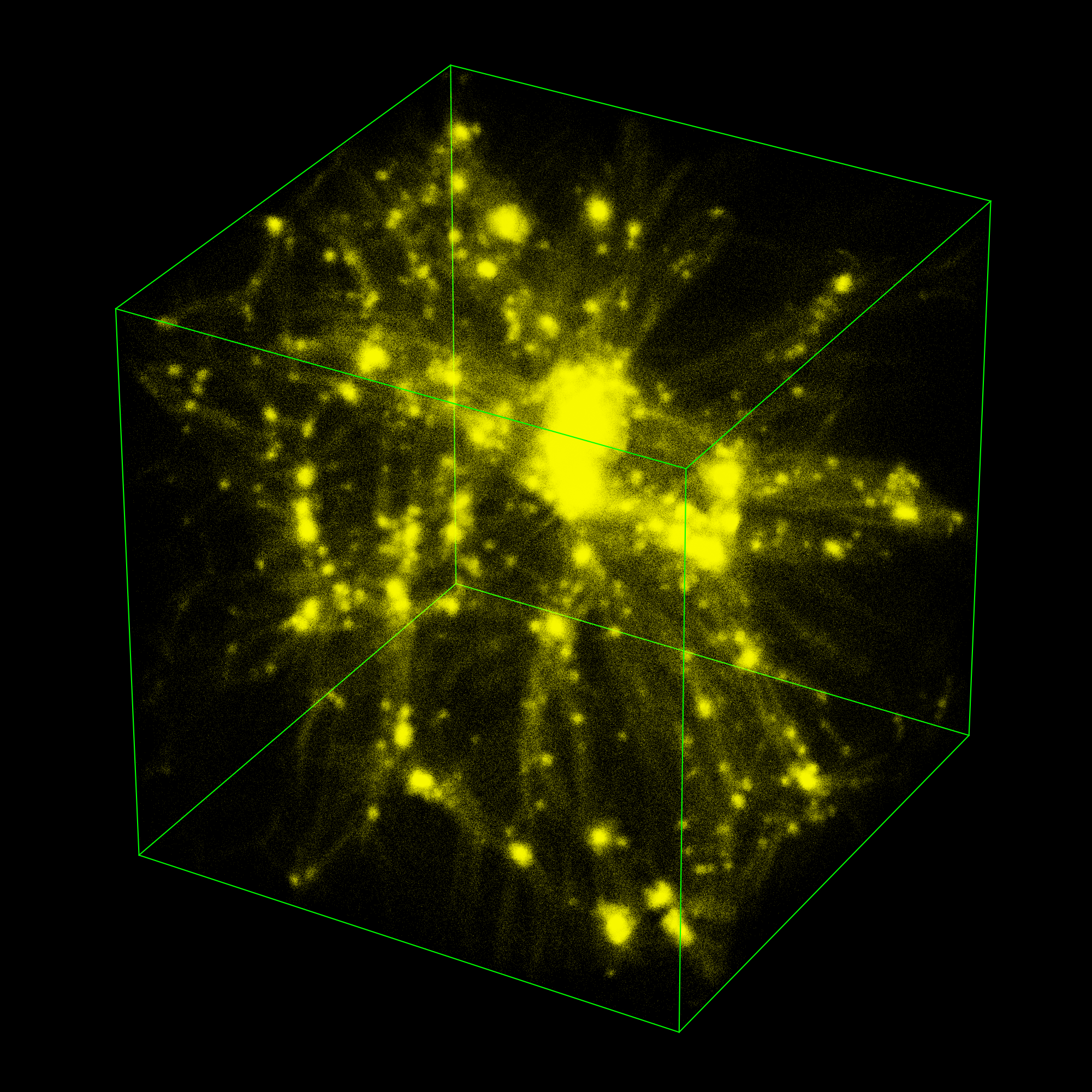} {\hskip 10pt}
\includegraphics[width=0.40\textwidth]{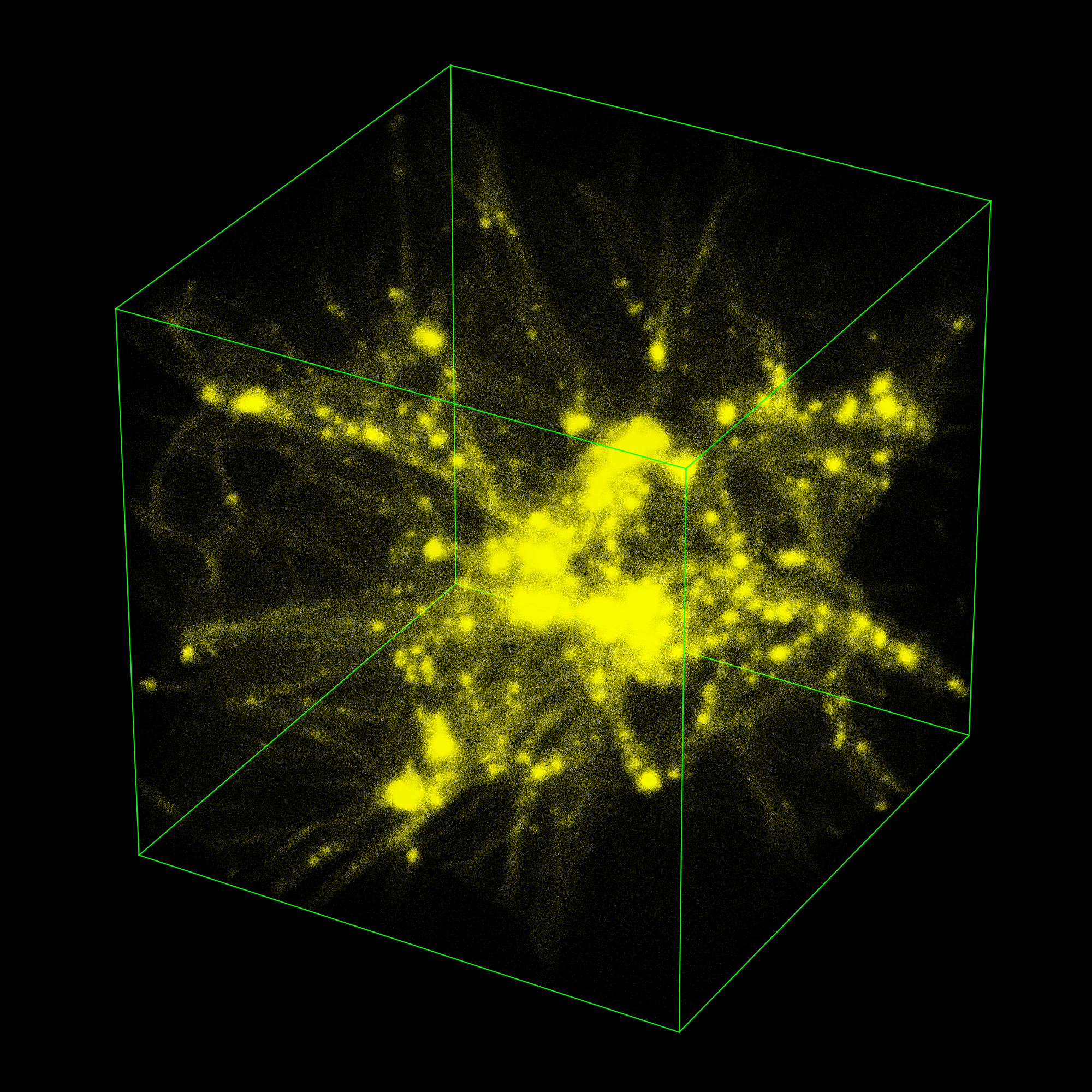}
}
\caption{Three-dimensional views of the four cosmic structures at $z=0$, selected from WIGEON. The views are for a void (top-left), a sheet (top-right), a filament (bottom-left), and a cluster (bottom-right), respectively. The length of each side of the box is $14.06\mpch$.}
\label{fig:wigeon4z00}
\end{figure*}

\section{Simulations: WIGEON and IllustrisTNG}
\label{sec:data}

In this work, we use two suites of simulation data, WIGEON and IllustrisTNG, for the computations. In the previous work, we have already made use of the WIGEON data \citep{Yang2020}. The remarkable characteristic of WIGEON simulation is that it employs the WENO, a positivity-preserving  finite-difference scheme, for the hydro solver of cosmic baryon fluid \citep{Zhu2013, Feng2004}. Due to its hydro-solver's five-order accuracy, WIGEON is more effective to capture turbulent structures and shock-waves in the baryon fluid. At $z = 11$, a uniform ultraviolet background (UVB) is turned on to mimic the re-ionization process. The radiative cooling and heating processes are modeled following the approach of \citet{Theuns1998}, with a primordial chemical composition, i.e. ${\rm X} = 0.76, {\rm Y} = 0.24$. As analyzed by \citet{Kang2007} and \citet{Iapichino2011}, the processes such as UVB and radiative cooling should not significantly affect the spatial distribution of cosmic baryonic gas in regions of turbulent flows. Actually, we can roughly estimate the effects of UVB heating as follows. Since the UVB heating can reionize and maintain the ionization state of hydrogen with radiative cooling at low redshifts, its temperature can be roughly estimated as $T\sim13.6{\rm eV} =1.6\times10^5$K, which is one order of magnitude smaller than the effective temperature $10^6$K of the turbulent pressure \citep{Zhu2010}. As mentioned in Section~\ref{sec:intro}, \citet{Zhu2010} show that the cosmic baryons are in the fully developed turbulence on scales $<3\mpch$. Hence, we can neglect the impact of UVB heating on the cosmic baryons in regions of turbulent flows with scales $<3\mpch$. In the simulations, we do not take into account these processes such as stellar formation and evolution, metal enrichment, SN and AGN feedback. By excluding these feedback processes, we expect that the effects of turbulent heating of IGM can be revealed to the most extent.

\begin{figure*}
\centerline{\includegraphics[width=0.80\textwidth]{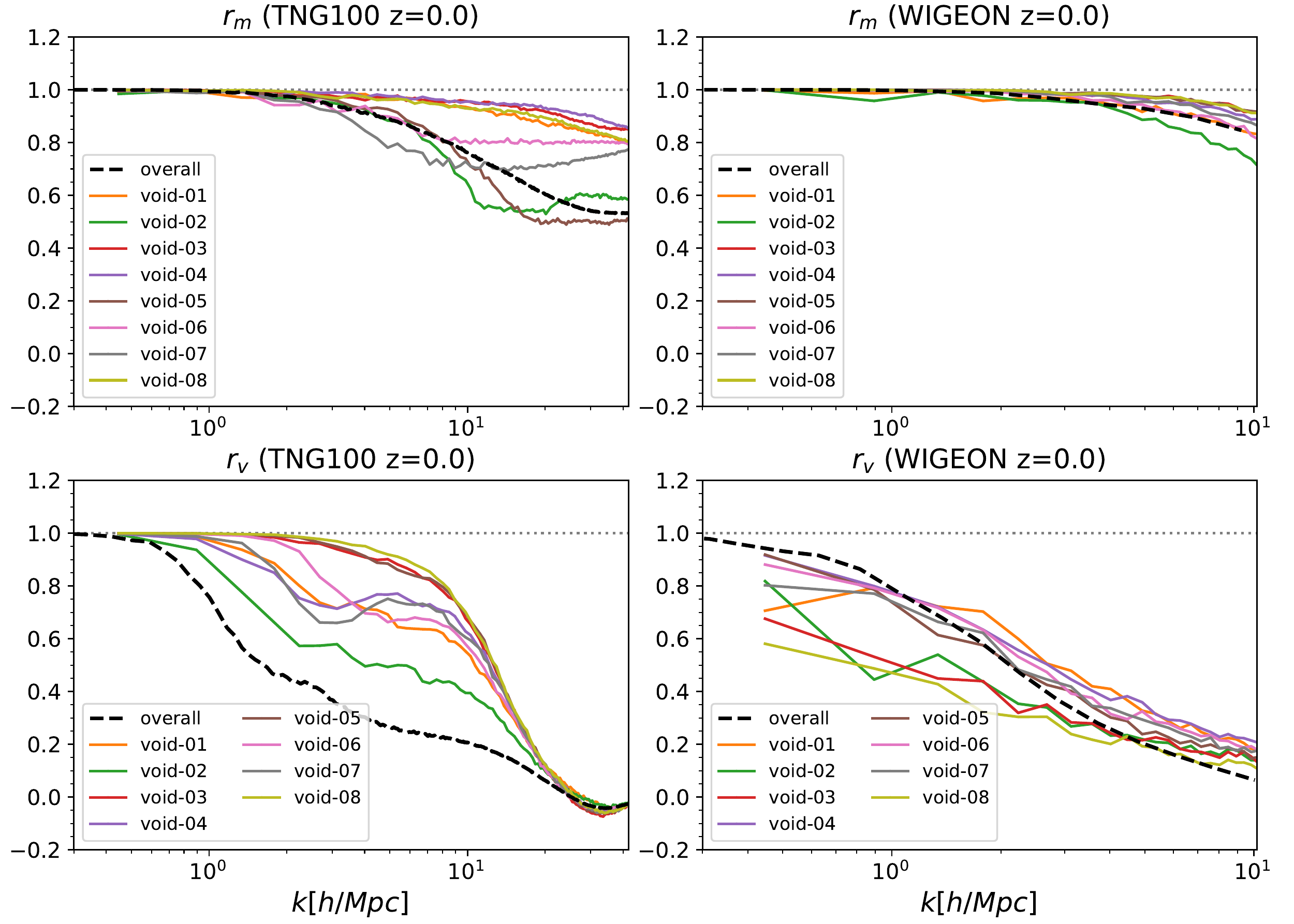}}
\caption{$r_{\rm m}$ and $r_{\rm v}$ of voids at $z=0$, data from TNG (left) and WIGEON (right). The eight thin lines indicate the eight voids. The thick black lines indicate the overall results of TNG100 or WIGEON, which are the same as in Fig.~\ref{fig:rmrv}.}
\label{fig:rmv-void-z00}
\end{figure*}

\begin{figure*}
\centerline{\includegraphics[width=0.80\textwidth]{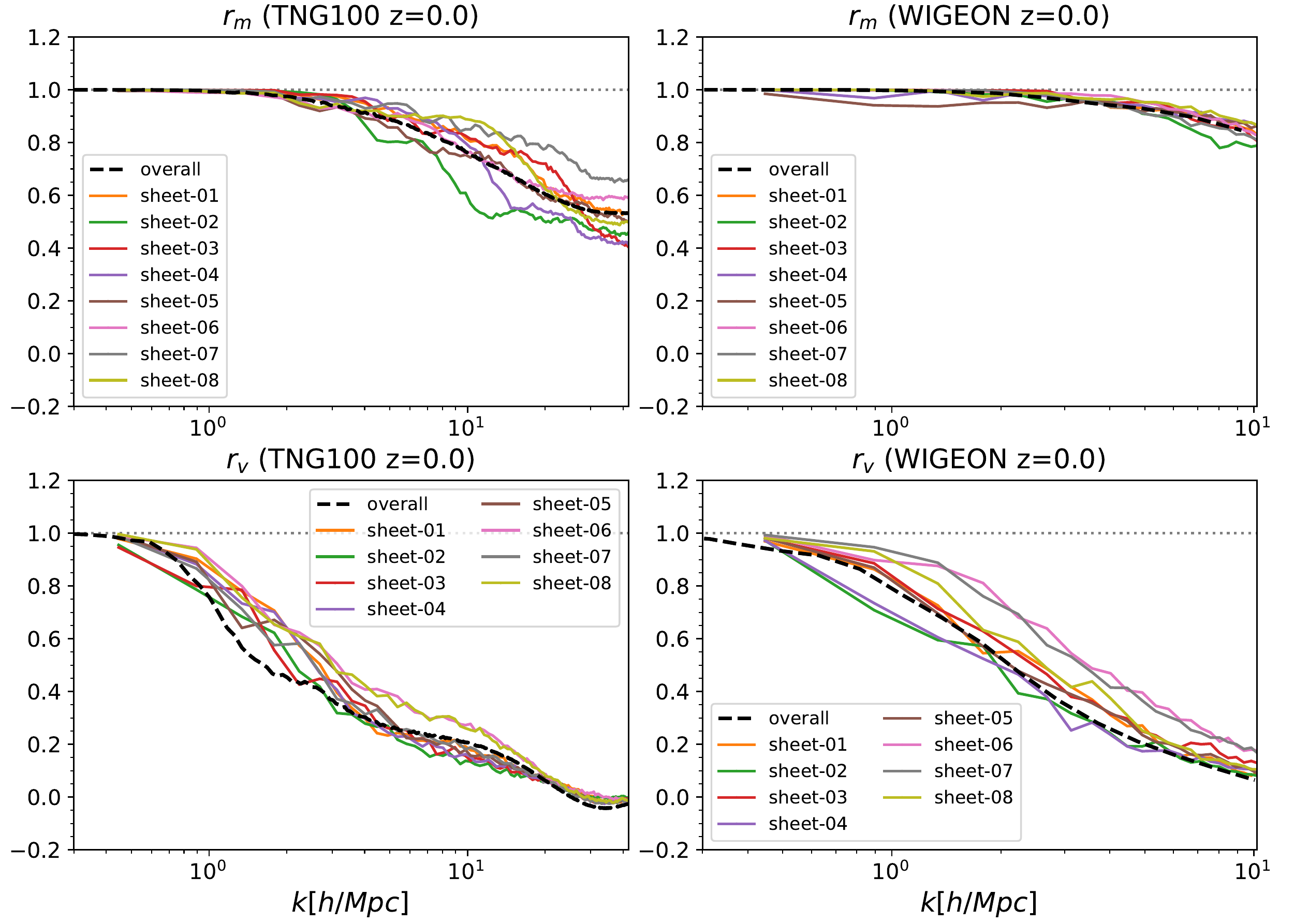}}
\caption{$r_{\rm m}$ and $r_{\rm v}$ for sheets at $z=0$, data from TNG (left) and WIGEON (right). The eight thin lines indicate the eight sheets. The thick black lines indicate the overall results of TNG100 or WIGEON, which are the same as in Fig.~\ref{fig:rmrv}.}
\label{fig:rmv-sheet-z00}
\end{figure*}

\begin{figure*}
\centerline{\includegraphics[width=0.80\textwidth]{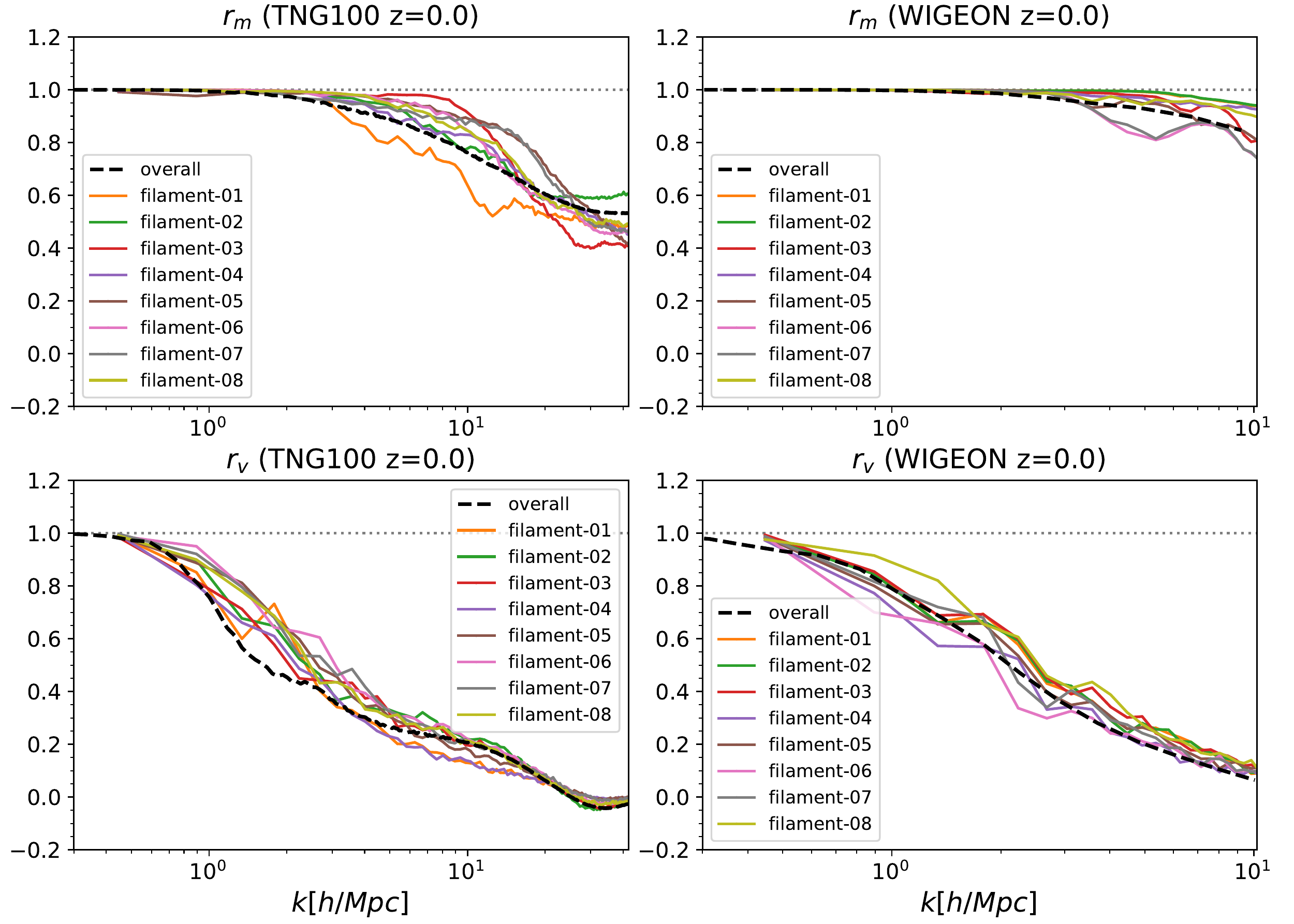}}
\caption{$r_{\rm m}$ and $r_{\rm v}$ for filaments at $z=0$, data from TNG (left) and WIGEON (right). The eight thin lines indicate the eight filaments. The thick black lines indicate the overall results of TNG100 or WIGEON, which are the same as in Fig.~\ref{fig:rmrv}.}
\label{fig:rmv-filament-z00}
\end{figure*}

\begin{figure*}
\centerline{\includegraphics[width=0.80\textwidth]{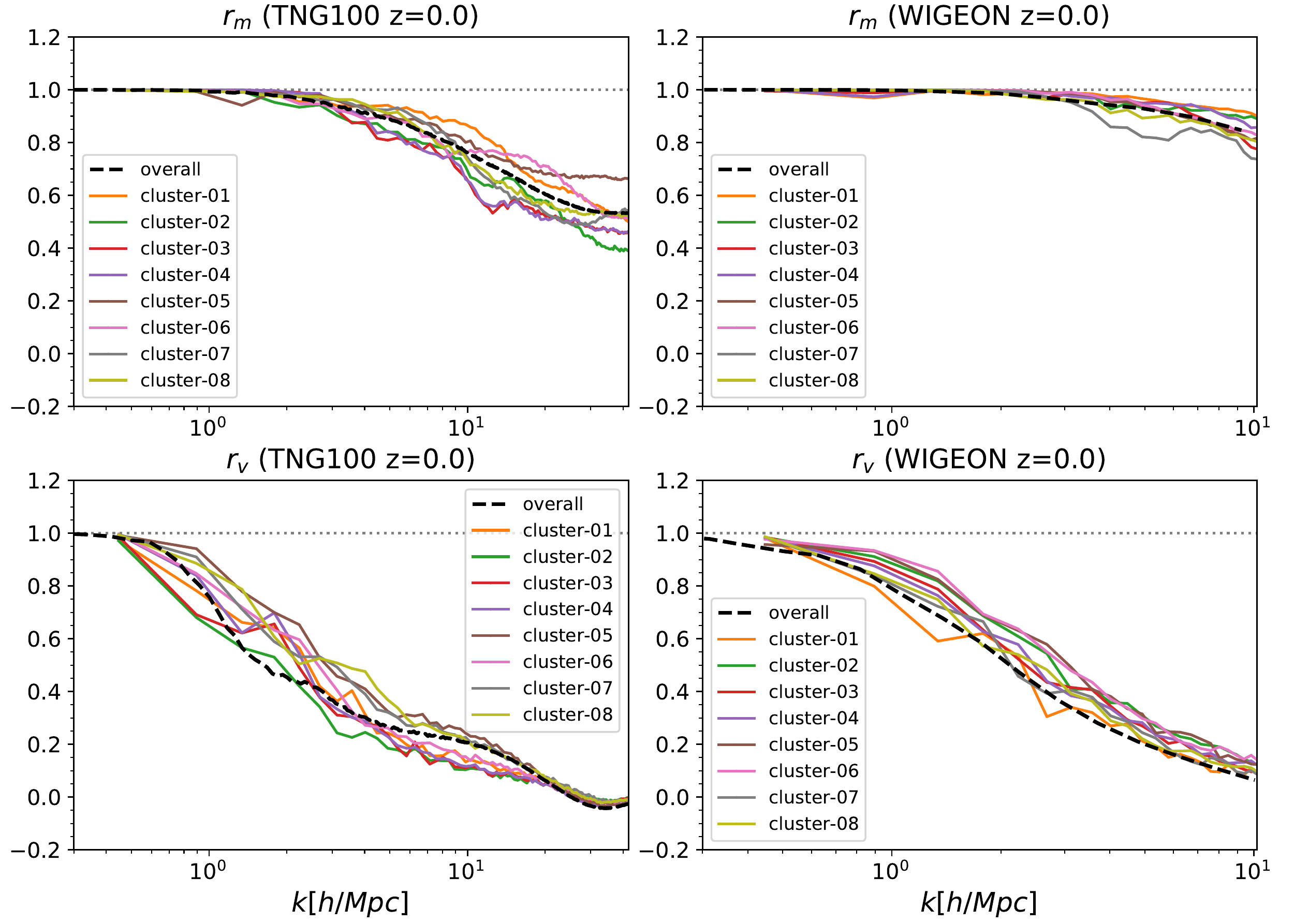}}
\caption{$r_{\rm m}$ and $r_{\rm v}$ for clusters at $z=0$, data from TNG (left) and WIGEON (right). The eight thin lines indicate the eight clusters. The thick black lines indicate the overall results of TNG100 or WIGEON, which are the same as in Fig.~\ref{fig:rmrv}.}
\label{fig:rmv-cluster-z00}
\end{figure*}

\begin{figure*}
\centerline{\includegraphics[width=0.80\textwidth]{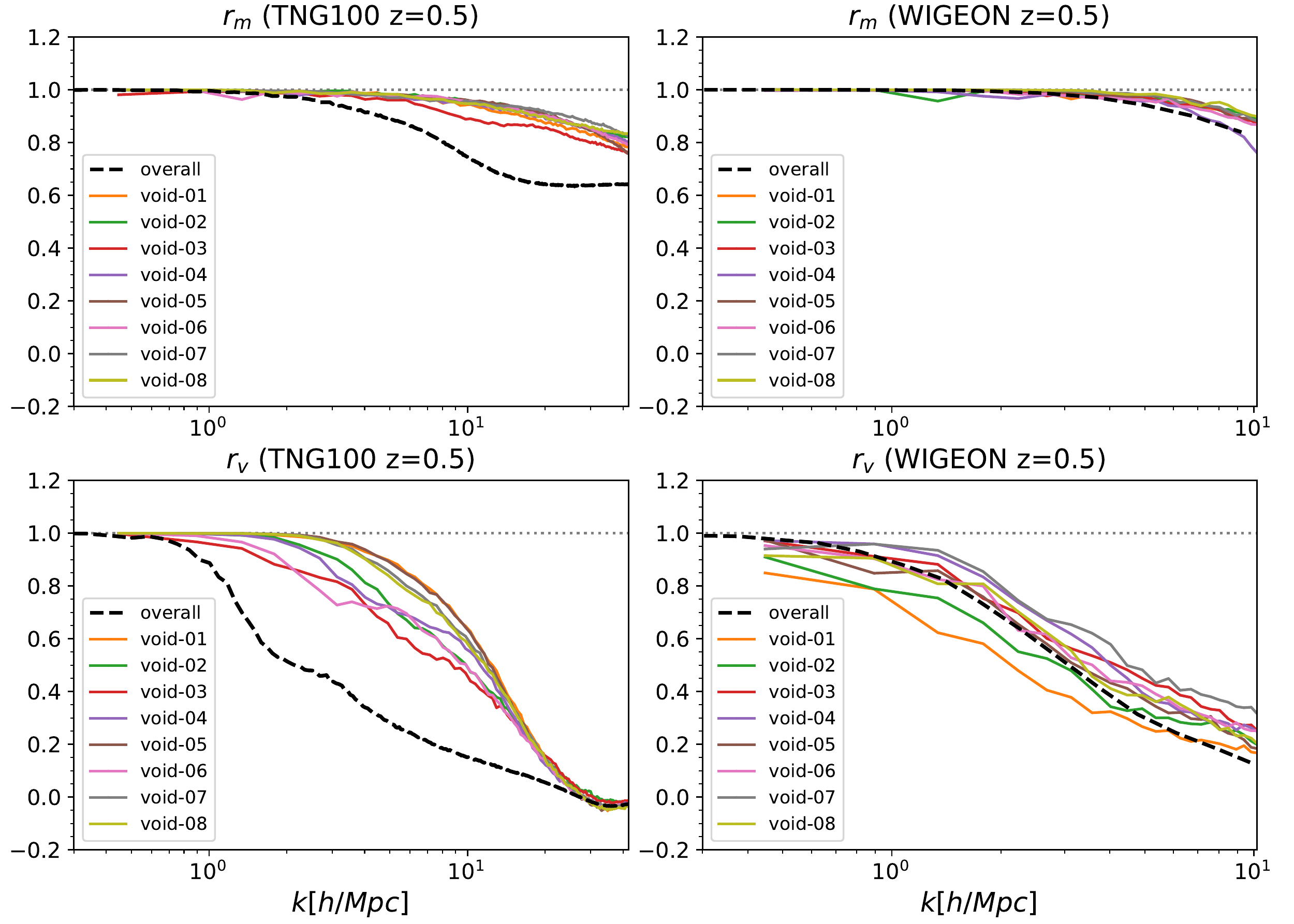}}
\caption{$r_{\rm m}$ and $r_{\rm v}$ of voids at $z=0.5$, data from TNG (left) and WIGEON (right). The eight thin lines indicate the eight voids. The thick black lines indicate the overall results of TNG100 or WIGEON, which are the same as in Fig.~\ref{fig:rmrv}.}
\label{fig:rmv-void-z05}
\end{figure*}

We also use the IllustrisTNG simulation data \citep{Pillepich2018a, Springel2018, Marinacci2018, Nelson2018, Naiman2018, Nelson2019}, from which we select the sample {\tt IllustrisTNG100-1} (TNG, hereafter), whose simulation box length is $75{\rm Mpc}/h$. IllustrisTNG is a suite of large volume, gravo-magnetohydrodynamical cosmological simulations, run with the moving-mesh code {\tt AREPO} \citep{Springel2010}. Besides gravity computation, all of the IllustrisTNG runs take into account the additional physical ingredients as follows \citep{Pillepich2018b, Nelson2019}: (1) stellar formation and evolution, (2) associated metal enrichment and mass loss, (3) primordial and metal-line radiative cooling, (4) pressurization of the interstellar medium (ISM) from unresolved SN, (5) stellar feedback, i.e. galactic-scale outflows with a kinetic wind driven by SN or asymptotic giant branch (AGB) stars, (6) formation and growth of supermassive black holes, with associated AGN feedback, i.e. releasing energy in the high-accretion-rate quasar mode and low-accretion-rate kinetic wind mode, and (7) influences of magnetic fields.

AREPO employs the second-order accurate finite volume Godunov-type scheme to solve hydrodynamical equations, which is formulated on an unstructured moving-mesh. After comparing the performance of smoothed particle hydrodynamics, \citet{Bauer2012} claimed that AREPO can better describe both supersonic and subsonic turbulence in the fluid, in that it yields Kolmogorov-like universal scaling laws for the power spectra of the density, velocity, and vorticity, which are consistent with expectations from the isotropic fully developed turbulence. Hence the TNG simulations include not only these processes such as stellar formation and evolution, metal enrichment, SN, and AGN feedback but also turbulence effects. All these physical processes will affect the spatial distribution of the cosmic baryon fluid.

\section{Results}
\label{sec:result}

\begin{figure*}
\centerline{\includegraphics[width=0.80\textwidth]{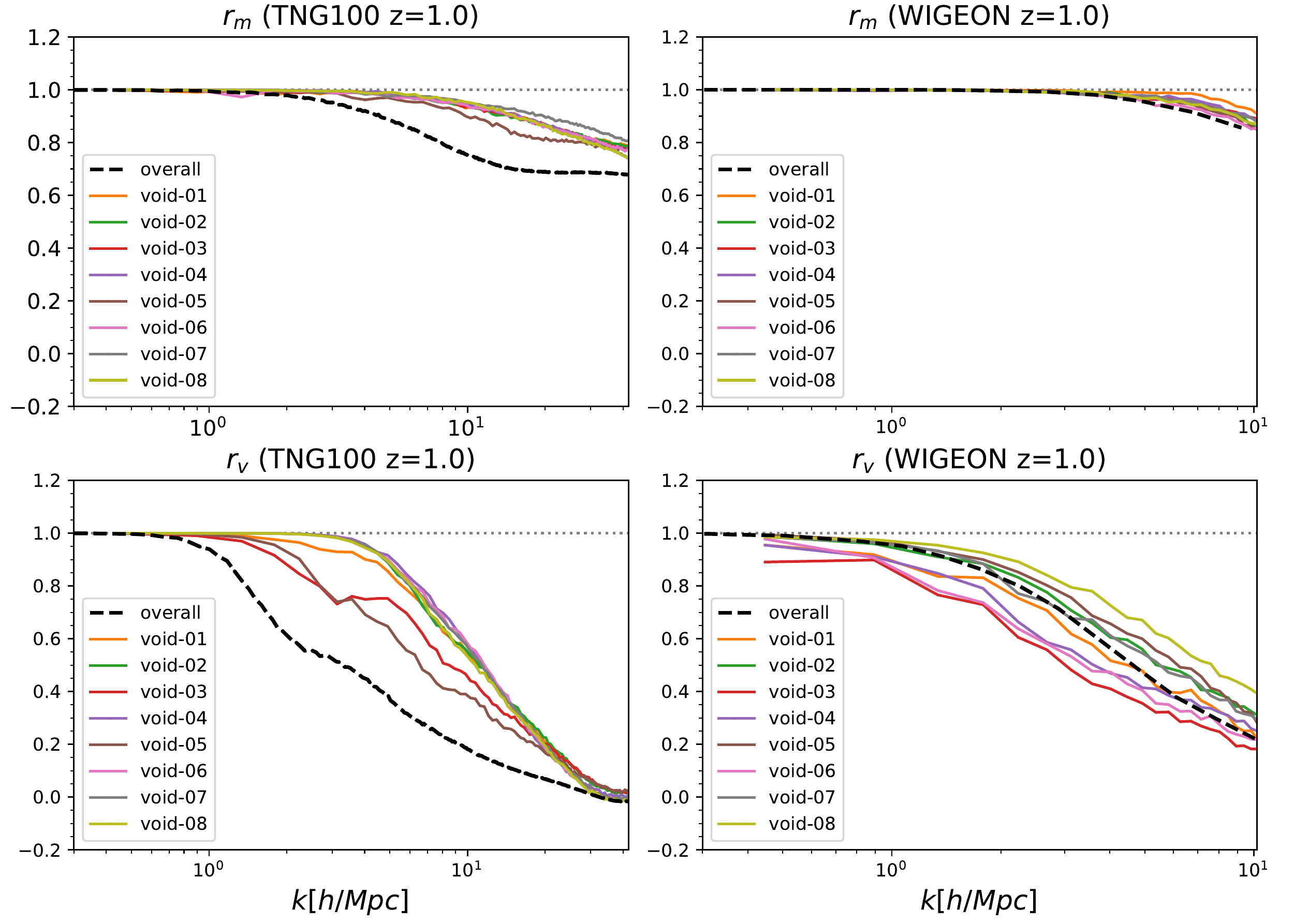}}
\caption{$r_{\rm m}$ and $r_{\rm v}$ of voids at $z=1.0$, data from TNG (left) and WIGEON (right). The eight thin lines indicate the eight voids. The thick black lines indicate the overall results of TNG100 or WIGEON, which are the same as in Fig.~\ref{fig:rmrv}.}
\label{fig:rmv-void-z10}
\end{figure*}

\subsection{Scale-dependent Deviation}
\label{sec:sclbias}

Applying the fast Fourier transform (FFT) to the WIGEON and TNG100 data, we compute the Fourier modes $\delta_{\rm dm}(\vk)$ and $\delta_{\rm b}(\vk)$ of density field, and $\vv_{\rm dm}(\vk)$ and $\vv_{\rm b}(\vk)$ of velocity field, and then compute the two correlation functions according to equation~(\ref{eq:rmvk}). From Fig.~\ref{fig:rmrv}, it can be seen that the results based on the two suites of simulation data are similar, which can be summarized as follows.

(i) Both the correlation functions ($r_{\rm m}$ and $r_{\rm v}$) approach one when $k$ goes to zero and decrease with $k$, which indicates that at increasingly large scales, the deviations between dark matter and baryons are vanishing, while at smaller and smaller scales, the deviations become gradually prominent.

(ii) All the $r_{\rm v}$ correlations are more significant than $r_{\rm m}$'s, especially at large $k$'s, which means that the deviation of velocity between dark matter and baryons is more remarkable than the deviation of density at increasingly smaller scales.

However, small differences do exist between the two suites of simulation data. From Fig.~\ref{fig:rmrv} we see that all the $r_{\rm m}$ correlations of TNG are larger than those of WIGEON at all scales, and this situation is not changed by redshift evolution. For the $r_{\rm v}$ correlations, situations are complicated. It can be seen that for $z=1$, $r_{\rm v}$ of TNG is larger than that of WIGEON at almost all valid scales; while for $z=0$, $r_{\rm v}$ of TNG is larger at $1<k<3\hmpc$, but smaller at $k>3\hmpc$ than that of WIGEON.

For such behavior of $r_{\rm v}$ correlations for WIGEON data, we refer the reader to Section~4.4 of \citet{Yang2020} for detailed analyses. Reiterating briefly here, the reasons should be as follows: (1) our hydro-solver is the WENO finite-difference scheme of five-order accuracy, which can compute the velocity of baryonic gas much accurately, and hence is much effective to capture turbulence and shockwave structures in IGM or ICM; (2) according to Helmholtz–Hodge decomposition \citep[cf.][]{Arfken2005}, a vector field \vv, can be decomposed into the divergence (or longitudinal) and the curl (or transverse) part. The divergence part of baryonic velocity fields tends to prevent baryons from falling into centres of gravitational potential wells, so that the spatial distribution of baryonic matter is more extended at later times than that of dark matter; (3) the curl part of velocity does not affect the time evolution of baryonic matter, and tends to increase at a faster pace than the divergence part with increasing time.

We briefly explain the similarities and small differences of the results between WIGEON and TNG simulations. As indicated by equation~(\ref{eq:rmest}), large scatters of the argument of the bias function $b(\vk)$ will lead to small $r_{\rm m}$ or $r_{\rm v}$ correlation coefficients. As mentioned in Section~\ref{sec:data}, we only consider turbulence effects in WIGEON simulations; while, besides turbulence, there are also strong SN or AGN feedback processes included in TNG data. The similarities of the $r_{\rm m}$ and $r_{\rm v}$ coefficients between WIGEON and TNG, imply that turbulence effects and SN/AGN feedback processes are somewhat degenerate, i.e. they have the same or similar effects to separate baryons from dark matter in the spatial distribution. While the differences of the $r_{\rm m}$ and $r_{\rm v}$ between WIGEON and TNG suggest that turbulence effects and SN/AGN feedback take into effect in different ways. After all, they are indeed different physical processes.

\subsection{Environment-dependent Deviation}
\label{sec:envbias}

The cosmic structures of the cosmic web are roughly classified into four categories, namely, voids, sheets (or walls), filaments, and clusters (or knots). We use the identification scheme by \citet{Hahn2007}, which identifies structures on the basis of the eigenvalues $\lambda_1$, $\lambda_2$, $\lambda_3$ of the tidal tensor. The tidal tensor is defined as the Hessian matrix of the re-scaled peculiar gravitational potential $\phi$,
\begin{equation}
\label{eq:ttensor}
T_{ij}=\frac{\partial^2\phi}{\partial x_i\partial x_j},
\end{equation}
where $i, j$=1, 2, and 3 denote the components in the three axes, and $x_i$, $x_j$ denote the comoving coordinates. The peculiar gravitational potential is re-scaled by $4\pi G\bar{\rho}(t)$, and obeys $\nabla^2 \phi = \delta$, where $\bar{\rho}(t)$ is the cosmic mean density of dark matter and $\delta = (\rho - \bar {\rho}) /\bar{\rho}$ is the overdensity field of dark matter. As in \citet{Zhu2017}, we count the number of eigenvalues above some threshold $\lambda_{\rm th}$ at each grid cell. A cell is assigned a value of 3 if the three $\lambda$'s are larger than $\lambda_{\rm th}$, and this cell is marked as a cluster. Similarly, by the same rule, a cell with the value of 2, 1, or 0 is marked as filament, sheet, or void, respectively.

In Fig.~\ref{fig:lambda-th}, we show the two-dimensional slice images of the cosmic web produced by TNG100 and WIGEON data at $z=0$, to compare with patterns of the spatial distribution of dark matter structures identified with different $\lambda_{\rm th}$. By visual comparison, we choose $\lambda_{\rm th}=0.5$ as the best value for both TNG and WIGEON data throughout the whole work. $\lambda_{\rm th}=0.5$ is also the best value adopted for other redshifts.

From both TNG and WIGEON data, we pick out the sub-samples of eight voids, eight sheets, eight filaments, and eight clusters in cubic regions, respectively. The box length of all the sub-samples is $14.06\mpch$. We manage to select such regions that each box can accommodate only one specific structure. However, we emphasize that it's hard or even impossible to select neatly only one specific structure in a cubic region without contamination by other kinds of structures. In Figs.~\ref{fig:tng4z00} and \ref{fig:wigeon4z00}, we show the three-dimensional images of $z=0$ for only one of the eight sub-samples of the four structures of TNG and WIGEON data, respectively.

In Figs.~\ref{fig:rmv-void-z00}, \ref{fig:rmv-sheet-z00}, \ref{fig:rmv-filament-z00}, and \ref{fig:rmv-cluster-z00}, we show $r_{\rm m}$ and $r_{\rm v}$ correlations of $z=0$ of the four structures for both TNG and WIGEON data. We see that almost all the results of the four structures fall around the corresponding overall results, with only one exception. From Fig.~\ref{fig:rmv-void-z00}, we see that $r_{\rm v}$ correlation of the TNG overall result is obviously larger than those of the eight TNG voids at almost all scales\footnote{According to the definitions of $r_{\rm m}$ and $r_{\rm v}$ in equation~(\ref{eq:rmvk}), the more deviation of $r_{\rm m}$ and $r_{\rm v}$ from one, the larger for the correlations, and vice versa.}. While the overall WIGEON correlation at $k<3\hmpc$ is smaller than but at $k>3\hmpc$ larger than those of the eight WIGEON voids.

At high redshifts, conclusions are similar to those at $z=0$, but $r_{\rm m}$ and $r_{\rm v}$ correlations become weaker as increasing redshifts. Remarkable differences also exist for TNG voids. In Figs.~\ref{fig:rmv-void-z05} and \ref{fig:rmv-void-z10}, we show the results for voids at $z=0.5$ and $1.0$. It can be seen that both $r_{\rm m}$ and $r_{\rm v}$ correlations of the TNG overall results are larger than those of the eight TNG voids, as is not observed in WIGEON data.

As addressed in Section~\ref{sec:tbasis}, $r_{\rm m}$ and $r_{\rm v}$ completely represent the non-linear spatial deviations between dark matter and baryons, and differences in the correlations between TNG and WIGEON data reflect the differences in the underlying physics and dynamics considered in the two simulations.

\begin{figure}
\centerline{\includegraphics[width=0.50\textwidth]{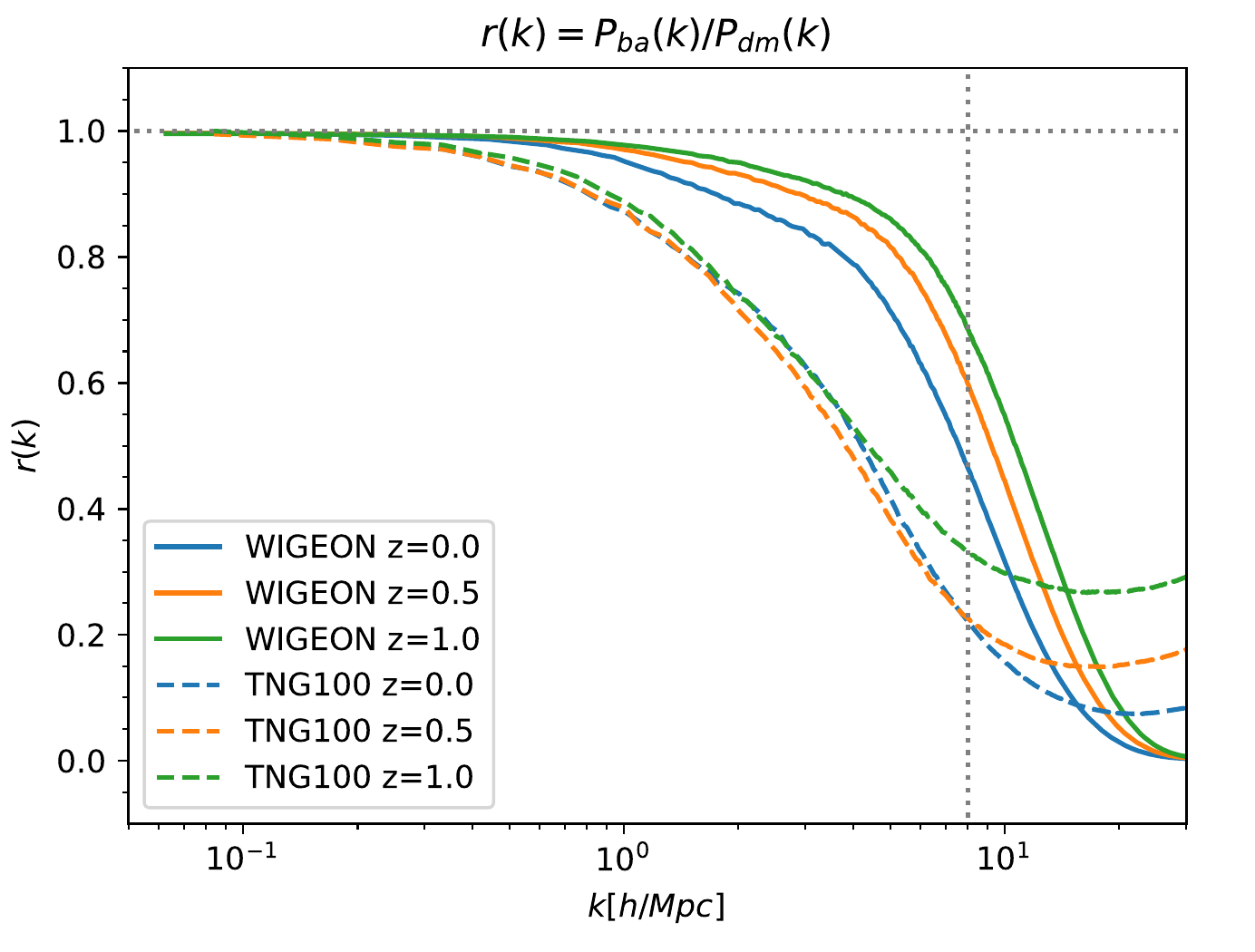}}
\caption{The ratio of density power spectrum between baryons and dark matter. For baryons, there is only the gaseous component in WIGEON simulation. While in TNG100 simulation, besides the gaseous components, there are also stars and black holes, such that the resulted gaseous ratios of TNG100 is slightly smaller than one when $k\rightarrow0$. In order to give correct comparison with WIGEON data, we artificially multiply the ratios of TNG100 by a factor to normalize the ratios to one at $k\rightarrow0$. The factor is 1.05 at $z=0$, 1.06 at $z=0.5$, and 1.07 at $z=1$, respectively. We draw a vertical dashed line at $k=8\hmpc$ to indicating the valid scale range for WIGEON is $k<8\hmpc$, which is analyzed in \citet{Yang2020}.}
\label{fig:powratio}
\end{figure}

\begin{figure}
\centerline{\includegraphics[width=0.50\textwidth]{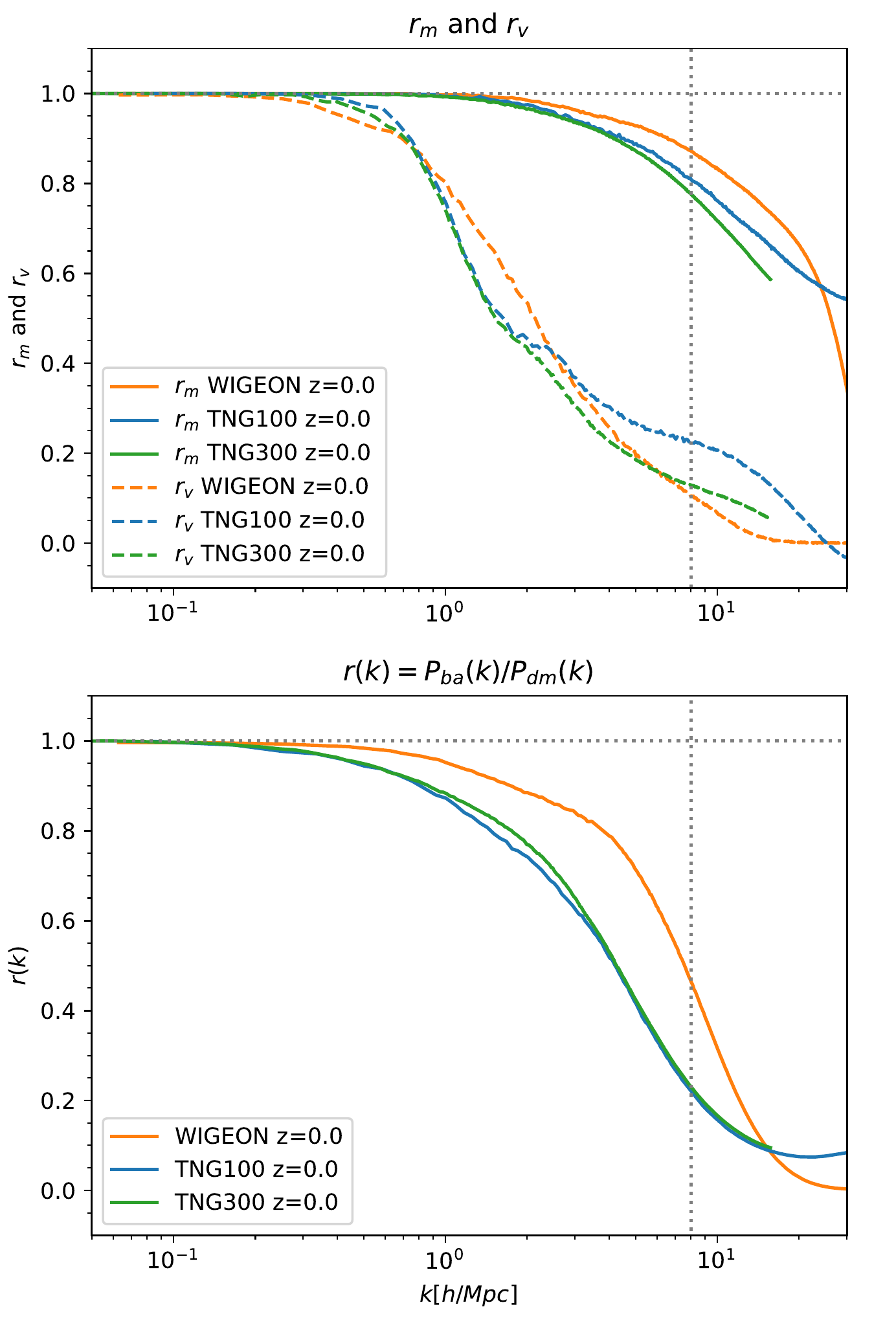}}
\caption{$r_{\rm m}$ and $r_{\rm v}$ (upper panel), and the ratio of density power spectrum (lower panel) for TNG100-1, TNG300-1 and WIGEON at $z=0$. As in Figure~\ref{fig:powratio}, we also multiply the ratio of TNG300 by a factor, 1.03, to normalize it at $k\rightarrow0$ to one.}
\label{fig:tng_comp}
\end{figure}

\subsection{Ratio of Power Spectrum}
\label{sec:pratio}

\begin{table}
\vspace{5pt}
\begin{center}
\begin{tabular*}{0.45\textwidth}{c|ccccc}
\hline\hline
$k (\hmpc)$ & $r_{\rm WIG}$ & $\Delta r_{\rm WIG}$ & $r_{\rm TNG}$ & $\Delta r_{\rm TNG}$ &
$R_{\rm WT}$ \\
\hline
2.0 & 0.884 & 0.116  & 0.741 & 0.259  & 0.45 \\
3.0 & 0.840 & 0.160  & 0.624 & 0.376  & 0.43 \\
4.0 & 0.788 & 0.212  & 0.517 & 0.483  & 0.44 \\
5.0 & 0.713 & 0.287  & 0.415 & 0.585  & 0.49 \\
6.0 & 0.629 & 0.371  & 0.331 & 0.669  & 0.55 \\
7.0 & 0.544 & 0.456  & 0.265 & 0.735  & 0.62 \\
8.0 & 0.466 & 0.534  & 0.223 & 0.777  & 0.69 \\
\hline
\end{tabular*}
\caption{$r_{\rm WIG}$ and $r_{\rm TNG}$ denote the ratio of power spectrum at $z=0$ for WIGEON and TNG, respectively, corresponding to the results in Fig.~\ref{fig:powratio}. $\Delta r_{\rm WIG} = 1 - r_{\rm WIG}$, $\Delta r_{\rm TNG} = 1 - r_{\rm TNG}$, and $R_{\rm WT} = \Delta r_{\rm WIG} / \Delta r_{\rm TNG}$.}
\end{center}
\label{tab1}
\end{table}

We define the ratio of density power spectrum between baryons and dark matter as $r(k)=P_{\rm ba}(k)/P_{\rm dm}(k)$, where $P_{\rm ba}(k)$ and $P_{\rm dm}(k)$ are matter density power spectrum of baryons and dark matter, respectively. In Fig.~\ref{fig:powratio}, we show the ratios for both WIGEON and TNG data at $z=0, 0.5, 1.0$. It can be seen that as scales become smaller and smaller, i.e. $k\rightarrow\infty$, the power spectra of baryons are increasingly suppressed for WIGEON simulations; while for TNG simulations, the suppression stops at $k=15-20\hmpc$, and due to star formation on small scales, the power spectrum ratios increase when $k>20\hmpc$. The suppression of power ratio for WIGEON is also redshift-dependent. From $z=1$ to $z=0$, the power ratio decreases from about 70\% to less than 50\% at $k=8\hmpc$. For TNG simulations, the suppression of power ratio is enhanced with decreasing redshifts in the scale range $k>4\hmpc$, but is nearly unchanged with redshifts in $k<4\hmpc$.

Besides turbulence, as mentioned in Section~\ref{sec:data}, TNG simulations also include the processes as stellar formation and evolution, metal enrichment, primordial and metal-line radiative cooling, stellar feedback by SN or AGB stars, formation and growth of supermassive black holes and associated AGN feedback, and magnetic fields. These physical processes are powerful mechanisms that are able to heat the gas in and around dark matter halos, and prevent the gas from being accreted and forming stars, or expel the gas directly from dark halos. It is these processes that suppress the power ratio in TNG simulations. However, unlike TNG simulations, we do not include these physical processes into WIGEON simulations. The major heating mechanism of the gas in WIGEON simulation is not those baryonic processes but turbulence. Turbulent heating can also have the consequence to suppress the power spectrum ratio between dark matter and baryons. In the following, we present a rough estimation of to what extent the power can be suppressed by turbulence.

From left to right, in the second column of Table~\ref{tab1}, we list the baryonic power ratio for WIGEON at $z=0$, $r_{\rm WIG}$, corresponding to the result in Fig.~\ref{fig:powratio}, and $\Delta r_{\rm WIG} = 1 - r_{\rm WIG}$  of the third column, indicates the suppression of power for WIGEON, and so on for $r_{\rm TNG}$ and $\Delta r_{\rm TNG} = 1 - r_{\rm TNG}$ of TNG simulations. We assume the power suppression of TNG $\Delta r_{\rm TNG}$ is the complete and correct reflection of all the physical effects, which can be treated as the norm for the total power suppression. Since we may neglect UVB heating when $k>2\hmpc$, corresponding to the scale $3\mpch$, as explained in Section~\ref{sec:data}, $\Delta r_{\rm WIG}$ reflects only the effect of turbulence. Then the ratio $R_{\rm WT}$ of the last column in the table, can be roughly regarded as the degree of power suppression by turbulence. For example, at $k=2\hmpc$, $R_{\rm WT}=0.45$, then the power suppression by turbulence can be regarded as 45\% of the total suppression, and the other processes except turbulence, such as stellar formation and evolution, metal enrichment, primordial and metal-line cooling, SN and AGN feedback, and so on, account for the left 55\% power suppression. Note that the percentage of power suppression by turbulence from 45\% at $k=2\hmpc$ gradually increases to 69\% at $k=8\hmpc$, indicating the impact of turbulence on the cosmic baryons are more significant on small scales. On scales of $k<2\hmpc$, the UVB heating should be the dominant heating mechanism of cosmic baryons.

Additionally, in Fig.~\ref{fig:tng_comp}, we compare $r_{\rm m}$, $r_{\rm v}$ and the power spectrum ratio $r(k)$ of TNG300-1 with those of TNG100-1. The simulation box length of TNG300 is $205\mpch$. We see that the results of TNG300 and TNG100 are nearly the same, and hence our conclusions are not affected by the box length of simulations.

\section{Summary and Conclusions}
\label{sec:concl}

In this work, by comparing the results derived from IllustrisTNG and WIGEON data, we check whether the conclusions derived in \citet{Yang2020} still hold for IllustrisTNG simulation. If there are some differences between them, what are the reasons for the differences? We summarize our findings and results as follows:

(1) As in \citet{Yang2020}, the bias function $b(\vk)$ in the Fourier space for TNG data is also a complex function, which means that its counterpart in real-space, $b({\bf x})$, is an asymmetric function. It would be interesting if one can present an analytic or semi-analytic form for $b(\vk)$ in equation~(\ref{eq:bias}), which can properly describe its dependence on the Fourier mode $\vk$ and its time evolution. This analytic/semi-analytic form may be useful for modelling of the spatial distribution of the baryonic matter.

(2) For both WIGEON and TNG data, both the correlation functions $r_{\rm m}$ and $r_{\rm v}$ approach one at $k$ goes to zero and decrease with increasing $k$, which indicates that on increasingly large scales, the spatial-distribution deviations between dark matter and baryons are vanishing, while on smaller and smaller scales, the deviations will be increasingly prominent.

(3) All the $r_{\rm v}$ correlations are more significant than $r_{\rm m}$'s, especially at large $k$'s, which means that the deviation of velocity between dark matter and baryons is more remarkable than the deviation of density on increasingly smaller scales.

(4) All the $r_{\rm m}$ correlations of TNG are larger than those of WIGEON at all scales, and this situation is not changed by redshift evolution. For the $r_{\rm v}$ correlations, situations are complicated. For $z=1$, $r_{\rm v}$ of TNG is larger than that of WIGEON at almost all valid scales; while for $z=0$, $r_{\rm v}$ of TNG is larger at $1<k<3\hmpc$, but smaller at $k>3\hmpc$ than that of WIGEON.

(5) For both TNG and WIGEON data, almost all $r_{\rm m}$ and $r_{\rm v}$ correlations of $z=0$ of the four structures, voids, sheets, filaments, and clusters, fall around the corresponding overall results. There is only one exception that $r_{\rm v}$ correlation of the TNG overall result is obviously larger than those of the eight TNG voids at almost all scales. While the overall WIGEON correlation at $k<3\hmpc$ is smaller than but at $k>3\hmpc$ larger than those of the eight WIGEON voids. At high redshifts, conclusions are similar to those of $z=0$, but $r_{\rm m}$ and $r_{\rm v}$ correlations become weaker as increasing redshifts.

(6) For the ratio of density power spectrum between baryons and dark matter, as scales become smaller and smaller, the power spectra for baryons are more and more suppressed for WIGEON simulations; while for TNG simulations, the suppression stops at $k=15-20\hmpc$, and due to star formation on small scales, the power spectrum ratios increase when $k>20\hmpc$. The suppression of power ratio for WIGEON is also redshift-dependent. From $z=1$ to $z=0$, the power ratio decreases from about 70\% to less than 50\% at $k=8\hmpc$. For TNG simulation, the suppression of power ratio is enhanced with decreasing redshifts in the scale range $k>4\hmpc$, but is nearly unchanged with redshifts in $k<4\hmpc$.

However, unlike TNG simulations, we do not take into account the processes such as stellar formation and evolution, metal enrichment, SN, and AGN feedback into WIGEON simulations. The heating mechanism of the gas in WIGEON simulation is not those baryonic processes but turbulence. Turbulent heating can also have the consequence to suppress the power ratio between baryons and dark matter. Regarding the power suppression for TNG simulations as the norm, the power suppression by turbulence for WIGEON simulations is roughly estimated to be 45\% at $k=2\hmpc$, and gradually increases to 69\% at $k=8\hmpc$, indicating the impact of turbulence on the cosmic baryons are more significant on small scales. On scales of $k<2\hmpc$, the UVB heating should be the dominant heating mechanism of cosmic baryons.

In addition to the studies by \citet{Fang2011}, \citet{Zhu2010}, \citet{Hep2006}, \citet{Zhu2013}, \citet{Zhu2015}, and \citet{Zhu2017}, an increasing amount of evidence has accumulated to support the turbulent heating mechanism of IGM. For example, \citet{Zhuravleva2014}, based on  deep X-ray data and a new data analysis method, find that turbulent heating is sufficient to compensate and balance the effects of radiative cooling locally at each core radius of Perseus and Virgo cluster of galaxies, indicating that turbulent heating to IGM is a necessary and significant heating mechanism. Another investigation by \citet{Nandakumar2020} shows evidence of large-scale energy cascade in the spiral galaxy NGC 5236, in which they find that the energy input scale to the interstellar medium turbulence is around $6{\rm kpc}$, driven by gravitational instability. This finding is also the evidence of turbulence occurring in IGM.

Besides those baryonic physics such as stellar formation and evolution, metal enrichment, SN, and AGN feedback, our results demonstrate that turbulent heating of IGM, is also of great importance to account for the motion and spatial distribution of cosmic baryons. In fact, \citet{Silk2010} show that AGN feedback may not be energetic enough to expel all the gas from the galaxy or even the halo \citep{Fabian2012}, and hence some form of extra heating mechanism such as turbulence is appealing. As mentioned in Section~\ref{sec:data}, AREPO uses the second-order accurate finite-volume Godunov-type scheme for the hydro-solver, while WIGEON employs the five-order accurate WENO finite-difference scheme. In view of the significance of turbulence in IGM, it is necessary for a comprehensive comparison of the performance of different hydro-solvers, which will be the target of our next work.

\section*{Acknowledgments}

We are very grateful for the referee's helpful and constructive comments and suggestions. The WIGEON simulations were run at Super-computing Center of the Chinese Academy of Sciences, and
SYSU. PH acknowledges the support by the Natural Science Foundation of Jilin Province, China (No. 20180101228JC), and by the National Science Foundation of China (No. 12047569, 11947415). We also acknowledge the use of the data from IllustrisTNG simulation for this work.

\section*{Data availability}

The data used in this paper are available from the correspondence author upon reasonable request.


\end{document}